\documentclass[%
reprint,
superscriptaddress,
amsmath,amssymb,
aps,
prb,
]{revtex4-2}

\usepackage{graphicx}% Include figure files
\usepackage{dcolumn}% Align table columns on decimal point
\usepackage{booktabs}% table tool
\usepackage{bm}% bold math
\usepackage{hyperref}% add hypertext capabilities
\usepackage{braket} % braket package
\usepackage{mathtools}
\usepackage{blkarray}
\usepackage{comment}
\hypersetup{
 colorlinks=true,
 linkcolor=blue,
 citecolor=blue,
 urlcolor=blue,
}

\begin{document}

%\preprint{APS/123-QED}

\title{\textbf{Paramagnetic electron-nuclear spin entanglement in \boldmath HoCo$_2$Zn$_{20}$} 
}% 

\author{Takafumi Kitazawa}
\email{kitazawa.takafumi@phys.sci.hokudai.ac.jp}
\affiliation{Advanced Science Research Center, Japan Atomic Energy Agency, Tokai, Ibaraki 319-1195, Japan}
\altaffiliation{Present address: Department of Physics, Faculty of Science, Hokkaido University, Sapporo 060-0810, Japan.}

\author{Yasuyuki Shimura}
\affiliation{Department of Quantum Matter, Graduate School of Advanced Science and Engineering, Hiroshima University, Higashi-Hiroshima 739-8530, Japan}

\author{Takahiro Onimaru}
\affiliation{Department of Quantum Matter, Graduate School of Advanced Science and Engineering, Hiroshima University, Higashi-Hiroshima 739-8530, Japan}

\author{Shun Tsuchida}
\affiliation{Graduate School of Science and Technology, Niigata University, Niigata 950-2181, Japan}

\author{Katsunori Kubo}
\affiliation{Advanced Science Research Center, Japan Atomic Energy Agency, Tokai, Ibaraki 319-1195, Japan}

\author{Yoshinori Haga}
\affiliation{Advanced Science Research Center, Japan Atomic Energy Agency, Tokai, Ibaraki 319-1195, Japan}

\author{Hironori Sakai}
\affiliation{Advanced Science Research Center, Japan Atomic Energy Agency, Tokai, Ibaraki 319-1195, Japan}

\author{Yoshifumi Tokiwa}
\affiliation{Advanced Science Research Center, Japan Atomic Energy Agency, Tokai, Ibaraki 319-1195, Japan}

\author{Shinsaku Kambe}
\affiliation{Advanced Science Research Center, Japan Atomic Energy Agency, Tokai, Ibaraki 319-1195, Japan}

\author{Yo Tokunaga}
\affiliation{Advanced Science Research Center, Japan Atomic Energy Agency, Tokai, Ibaraki 319-1195, Japan}

\date{\today}% It is always \today, today,
             %  but any date may be explicitly specified

\begin{abstract}
We investigated electron-nuclear spin entanglement in the paramagnetic ground state of the Ho-based cubic compound HoCo$_2$Zn$_{20}$.  
From analyses of magnetization and specific heat data, we determined the cubic crystalline electric field (CEF) parameters, the magnetic exchange constant, and the hyperfine coupling constant between the 4$f$ magnetic moment and the $^{165}$Ho nuclear spin. 
Our results show that the $\Gamma_5$ CEF ground state is split by the hyperfine coupling, with an energy width of 1.3~K at 0~T, and that the true paramagnetic ground state is a quasi-sextet arising primarily from entanglement between the $f$-electron effective spin $S = 1$ and the $^{165}$Ho nuclear spin $I = 7/2$. 
We further demonstrate that, depending on the CEF parameters, the paramagnetic ground state can switch to an electron-nuclear coupled dectet.
These findings underscore the importance of accurately identifying the electron-nuclear level scheme for understanding the low-temperature properties of rare-earth compounds containing spin-active nuclei. 
\end{abstract}

%\keywords{Suggested keywords}%Use showkeys class option if keyword
%display desired
\maketitle

%\tableofcontents

\section{\label{sec:introduction}Introduction}

In studies of rare-earth-based compounds, the energy-level scheme at rare-earth sites—particularly the nature of the ground state, and in some cases the first excited state—provides essential information for understanding exotic phenomena that emerge at low temperatures.  
When discussing the level scheme at rare-earth sites, we typically begin with the ground $J$ multiplet within the $LS$-coupling scheme, since the Coulomb interaction among the 4$f$ electrons is much stronger than their spin-orbit coupling~\cite{Santini1999}.  
In crystals, the ground $J$ multiplet is further split by the crystalline electric field (CEF).  
The CEF parameters have been determined using a variety of experimental probes, including specific heat, magnetization, the elastocaloric effect~\cite{Gati2023}, Raman scattering~\cite{Cai2024, Wang2024}, inelastic neutron scattering~\cite{Yamamoto2023a, Anand2023, Tsukagoshi2023, Ueta2024, Zheng2024, Yamauchi2025}, and x-ray spectroscopic techniques~\cite{Zhao2023, Amorese2023, Christovam2024}. 

Although most electronic properties of rare-earth-based compounds can be understood in terms of CEF splitting alone, it should be emphasized that, when a nuclear spin is present at the rare-earth site, the CEF states are no longer exact eigenstates even in zero magnetic field.  
This arises from the hyperfine coupling between the magnetic dipole moment of the $f$ electrons, $\bm{J}$, and the nuclear spin $\bm{I}$ at the rare-earth site, expressed as $A_\mathrm{HF}\,\bm{I}\cdot\bm{J}$, where $A_\mathrm{HF}$ is the hyperfine coupling constant.  
As a result, the CEF eigenlevels are split, and each eigenstate becomes a coupled electron–nuclear state.  
This splitting can often be neglected in magnetically ordered rare-earth-based compounds, since their ordering temperatures are typically much higher than the energy scale of the hyperfine interaction.  
In contrast, the hyperfine coupling becomes significant in certain exotic materials, such as spin ices~\cite{Dun2020, Gronemann2023} and quantum critical materials~\cite{Eisenlohr2021, Bitko1996, McKenzie2018, Libersky2021, Knapp2023, Knapp2025, Levitin2025, Banda2023}, where localized electrons remain in the paramagnetic state even at very low temperatures.  
In such systems, determining the paramagnetic level scheme while accounting for the hyperfine interaction is crucial for understanding quantum phenomena. 

An important material-dependent parameter in this context is the hyperfine coupling constant $A_\mathrm{HF}$.  
However, experimental techniques for determining $A_\mathrm{HF}$ at rare-earth sites have so far been limited to specific microscopic probes~\cite{Shakurov2005, Beckert2022}.  
Here, we demonstrate that specific heat can serve as an alternative probe for estimating $A_\mathrm{HF}$ and that the hyperfine electron–nuclear coupled wave function in the paramagnetic state can be extracted solely from macroscopic thermodynamic measurements. 

As a candidate material for investigating the paramagnetic electron-nuclear state, we focused on the holmium-based compound HoTr$_2$Zn$_{20}$ (Tr = transition metal) for the following three reasons.  
(1) The $^{165}$Ho isotope, which is 100\% naturally abundant, has the largest nuclear spin, $I = 7/2$, among the lanthanides excluding the nonmagnetic elements La and Lu~\cite{Rumble2024}.  
Therefore, a Schottky-like specific heat anomaly arising from large hyperfine splitting is expected to appear at relatively high temperatures.  
(2) The RTr$_2$Zn$_{20}$ (R = rare-earth element) family crystallizes in the CeCr$_2$Al$_{20}$-type structure, in which the nearest-neighbor distance between rare-earth ions is as large as 6~\AA~\cite{Nasch1997}, implying a very low ordering temperature due to weak dipole and/or multipole interactions between rare-earth ions.  
In fact, although the CEF ground state is not a singlet, $f$-electron ordering temperatures below 1~K have been reported in several RTr$_2$Zn$_{20}$ compounds with Co-group elements (Tr = Co, Rh, and Ir)~\cite{Onimaru2011, Onimaru2012, Yamane2017, Yamamoto2019}.  
(3) The rare-earth site has cubic point symmetry $T_d$, indicating the absence of a nuclear electric quadrupole interaction~\cite{Abragam1961}.  
Thus, the nuclear quadrupole effect does not need to be considered in calculations of thermodynamic quantities.

Among the HoTr$_2$Zn$_{20}$ compounds, HoCo$_2$Zn$_{20}$ is the only one for which no phase transition has been reported, although its physical properties above 1.85~K have been studied~\cite{Jia2008, Jia2009}. 
Thus, HoCo$_2$Zn$_{20}$ serves as a suitable candidate for examining the electron-nuclear state in the paramagnetic regime.  
In this paper, we first provide a brief overview of the experimental methods in Sec.~\ref{sec:method}, and then present macroscopic measurements down to approximately 0.3~K on single crystals of HoCo$_2$Zn$_{20}$ in Sec.~\ref{sec:results}.  
In Sec.~\ref{sec:analysis}, we introduce a model Hamiltonian for the Ho sites and determine its free parameters from the macroscopic experimental data.  
We then describe the electron-nuclear states at zero magnetic field in HoCo$_2$Zn$_{20}$.
Finally, in Sec.~\ref{sec:discussion}, we discuss the electron-nuclear level scheme in general cubic Ho compounds and the multichannel Kondo effect in realistic materials containing nuclear spins, and conclude the paper in Sec.~\ref{sec:conclusion}.

%
%
%
%%%%%%%%%%%%%%%%%%%%%%%%%%%%%%%%%%%%%%%
\section{\label{sec:method}Experimental methods}
%%%%%%%%%%%%%%%%%%%%%%%%%%%%%%%%%%%%%%%

Single crystals of HoCo$_2$Zn$_{20}$ and the nonmagnetic isostructural compound LuCo$_2$Zn$_{20}$ were grown using the Zn self-flux method.  
Pure elements were weighed in the nominal ratio of $\mathrm{R} : \mathrm{Co} : \mathrm{Zn} = 1 : 2 : 47$ ($\mathrm{R} = \mathrm{Ho}, \mathrm{Lu}$), following Ref.~\cite{Jia2009}, and placed into an aluminum crucible.  
Each batch of RCo$_2$Zn$_{20}$ was sealed in an evacuated quartz tube and heated to $1100~^\circ\mathrm{C}$.  
After maintaining this temperature for 3~h, the samples were slowly cooled to $700~^\circ\mathrm{C}$ at a rate of $-2~^\circ\mathrm{C}$/h.
The cubic lattice constant of HoCo$_2$Zn$_{20}$, determined by single-crystal x-ray diffraction using Mo K$_\alpha$ radiation and an R-AXIS RAPID diffractometer (Rigaku), was found to be $a = 14.0344(5)$~\AA.  
This value is consistent with $a = 14.028$~\AA, extracted from Fig.~2 in Ref.~\cite{Jia2009}, which shows the cubic lattice constants of RTr$_2$Zn$_{20}$ (R = rare earth) obtained by powder x-ray diffraction.

Magnetization measurements in the temperature range from 1.9 to 300~K were performed using a commercial dc superconducting quantum interference device magnetometer (MPMS XL7, Quantum Design).  
Additional magnetization measurements down to 0.27~K were carried out using a capacitive Faraday magnetometer~\cite{Sakakibara1994} mounted on a $^3$He refrigerator (Heliox $2^\mathrm{VL}$, Oxford Instruments).
Electrical resistivity and specific heat were measured using a physical property measurement system equipped with a $^3$He cooling option (PPMS DynaCool-9T, Quantum Design).  
Resistivity was measured using the standard four-probe method with the electrical transport option, while specific heat was measured using the thermal relaxation method.

%
%
%
%%%%%%%%%%%%%%%%%%%%%%%%%%%%%%%%%%%%%%%
\section{\label{sec:results} Experimental results}
%%%%%%%%%%%%%%%%%%%%%%%%%%%%%%%%%%%%%%%

\subsection{\label{subsec:overview} Overview of \boldmath HoCo$_2$Zn$_{20}$}

\begin{figure}[tb!]
\includegraphics[keepaspectratio, width=8.5cm, clip]{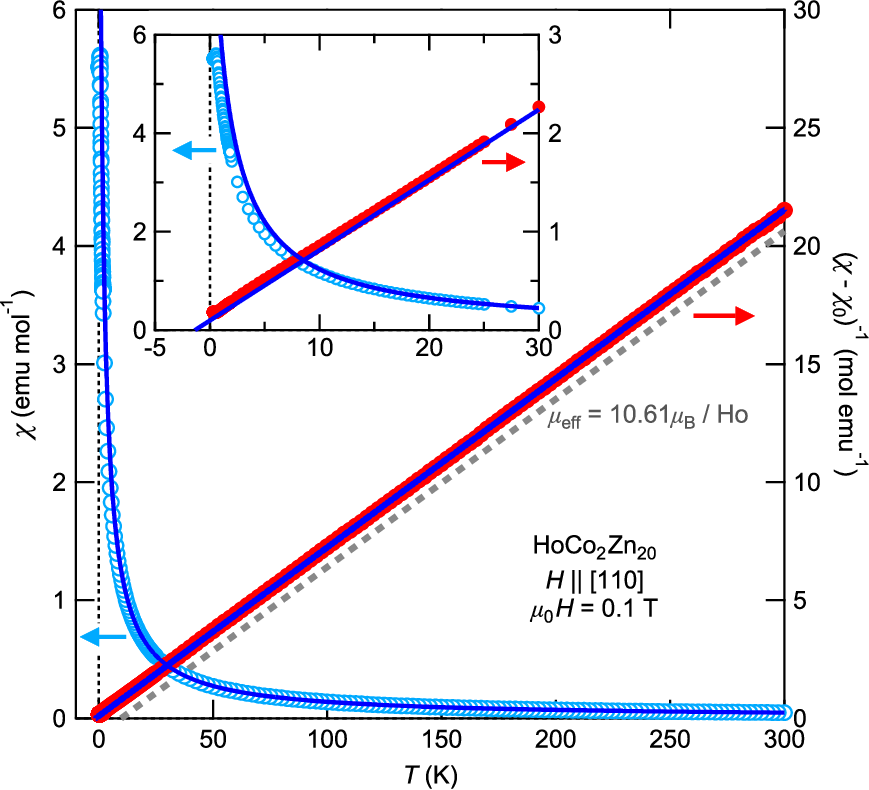}
\caption{\label{fig:chiT} 
Temperature dependence of the magnetization divided by the applied magnetic field $\mu_0 H = 0.1$~T, denoted as $\chi$ (left axis), and the inverse one, $(\chi - \chi_0)^{-1}$ (right axis), obtained after subtracting a temperature-independent term $\chi_0$, in HoCo$_2$Zn$_{20}$ for $H \parallel [110]$. 
Data for $H \parallel [100]$ and $[111]$ are presented in Fig.~\ref{fig:chiT_all}. 
The blue solid line represents a fit to the modified Curie-Weiss law, and the gray dashed line indicates the expected slope of the inverse susceptibility for free Ho$^{3+}$ ions. 
The inset shows an enlarged view of $\chi$ and $(\chi - \chi_0)^{-1}$ below 30~K.
}
\end{figure}

Figure~\ref{fig:chiT} shows the temperature dependence of the magnetization divided by the applied magnetic field $\mu_0 H = 0.1~\mathrm{T}$ in HoCo$_2$Zn$_{20}$. 
In this paper, we define $\chi$ as $M/H$ for $H = 0.1~\mathrm{T}$, and omit the data for $H \parallel [100]$ and $[111]$ from Fig.~\ref{fig:chiT}, since $\chi$ measured by MPMS is nearly isotropic as shown in Fig. \ref{fig:chiT_all} (Appendix \ref{app_sec:M_suppl}). 
The effective magnetic moment $\mu_\mathrm{eff}$ and the Curie-Weiss temperature $\theta_\mathrm{C}$ were evaluated by fitting the modified Curie-Weiss law,  
$\chi = W_\mathrm{C}/(T - \theta_\mathrm{C}) + \chi_0$, where $W_\mathrm{C}$ is the Curie constant and $\chi_0$ is the temperature-independent magnetic susceptibility.  
The fitting results in the temperature range 100--300~K are summarized in Table~\ref{tab:CW_fit_results}.  
The dashed line in Fig.~\ref{fig:chiT}, which runs parallel to the $(\chi - \chi_0)^{-1}$ data (red plot), clearly indicates that the $\mu_\mathrm{eff}$ of HoCo$_2$Zn$_{20}$ is very close to $10.61~\mu_\mathrm{B}/\mathrm{Ho}$, the value expected for the ground $J$ multiplet ($J = 8$) of the $4f^{10}$ configuration at Ho sites. 
This result confirms that HoCo$_2$Zn$_{20}$ hosts a localized magnetic moment at each Ho site. 
In addition to the value of $\mu_\mathrm{eff}$, the small magnitude of $\theta_\mathrm{C}$, which suggests a low magnetic transition temperature, is consistent with previously reported values ($\mu_\mathrm{eff} = 10.7~\mu_\mathrm{B}/\mathrm{Ho}$, $\theta_\mathrm{C} = 1.4~\mathrm{K}$)~\cite{Jia2009}.

%%%%%%%%%%%%%%%%%%%%%%%%%%%%%%%%%%%%%%%
%%%  Fig. 1
%%%%%%%%%%%%%%%%%%%%%%%%%%%%%%%%%%%%%%%
%%%%%%%%%%%%%%%%%%%%%%%%%%%%%%%%%%%%%%%
%%%
%%%%%%%%%%%%%%%%%%%%%%%%%%%%%%%%%%%%%%%

%%%%%%%%%%%%%%%%%%%%%%%%%%%%%%%%%%%%%%%
%%%  Table 1
%%%%%%%%%%%%%%%%%%%%%%%%%%%%%%%%%%%%%%%
\begin{table}[tb!]
\caption{\label{tab:CW_fit_results}
Effective magnetic moment $\mu_\mathrm{eff}$, Curie-Weiss temperature $\theta_\mathrm{C}$, and temperature-independent susceptibility $\chi_0$ in HoCo$_2$Zn$_{20}$, obtained from fits to the magnetic susceptibility $\chi(T)$ using the modified Curie-Weiss law in the temperature range 100--300~K.}
\catcode`!=\active \def!{\phantom{.}}
\catcode`?=\active \def?{\phantom{0}}
\begin{ruledtabular}
\begin{tabular}{rccc}
&
$\mu_\mathrm{eff} \ (\mu_\mathrm{B}/\mathrm{Ho})$ &
$\theta_\mathrm{C} \ (\mathrm{K})$ &
$\chi_0 \ (10^{-3} \ \mathrm{emu \ mol^{-1}})$ \\
\midrule
$H \ || \ [100]$ & 10.59(2) & $-$1.9(3) & 2.3(1) \\
$[110]$ & 10.58(2) & $-$1.4(3) & 2.1(1) \\
$[111]$ & 10.58(2) & $-$1.7(3) & 5.1(1) \\
\end{tabular}
\end{ruledtabular}
\end{table}
%%%%%%%%%%%%%%%%%%%%%%%%%%%%%%%%%%%%%%%
%%%
%%%%%%%%%%%%%%%%%%%%%%%%%%%%%%%%%%%%%%%

\begin{figure}[tb!]
\includegraphics[keepaspectratio, width=7.5cm, clip]{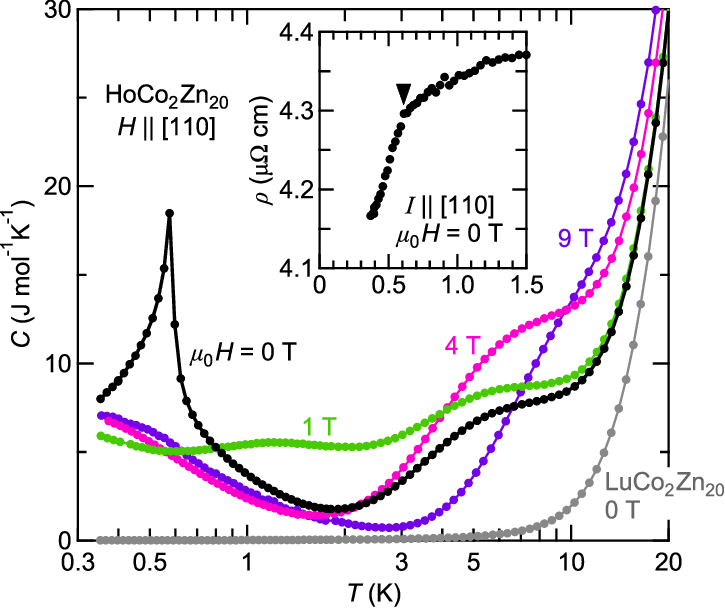}
\caption{\label{fig:CT_and_rhoT} 
Temperature dependence of the specific heat for HoCo$_2$Zn$_{20}$ at 0, 1, 4, and 9~T applied along the $[110]$ direction, and for LuCo$_2$Zn$_{20}$ at 0~T. The inset shows the electrical resistivity of HoCo$_2$Zn$_{20}$ at 0~T.
}
\end{figure}

Next, we present the temperature dependence of the specific heat $C(T)$ and resistivity $\rho(T)$ of HoCo$_2$Zn$_{20}$, shown in Fig.~\ref{fig:CT_and_rhoT}.  
At zero field, $C(T)$ and $\rho(T)$ exhibit a sharp peak and a kink, respectively, around 0.6~K, indicating a phase transition.  
Details of this transition are discussed in Sec.~\ref{subsec:phase_transition}. 

%%%%%%%%%%%%%%%%%%%%%%%%%%%%%%%%%%%%%%%
%%%  Fig. 2
%%%%%%%%%%%%%%%%%%%%%%%%%%%%%%%%%%%%%%%
%%%%%%%%%%%%%%%%%%%%%%%%%%%%%%%%%%%%%%%
%%%
%%%%%%%%%%%%%%%%%%%%%%%%%%%%%%%%%%%%%%%

In addition to the phase transition, several other features appear in $C(T)$.  
First, $C(T)$ at 0, 1, and 4~T shows a shoulderlike anomaly near 5~K.  
As revealed in the CEF level-scheme analysis of Sec.~\ref{subsec:cef_refinement}, this anomaly arises from thermal excitation between CEF levels.  
Second, $C(T)$ at 1~T does not fall below 5~$\mathrm{J \ mol^{-1} \ K^{-1}}$ even at the lowest measured temperature of 0.35~K, and $C(T)$ at 4 and 9~T exhibits an enhancement below about 2~K.  
These anomalies can be attributed to the nuclear specific heat from nuclear spins.  
As listed in Table~\ref{tab:list_nuclear_spin}, HoCo$_2$Zn$_{20}$ contains nuclear spins not only in Ho but also in Co and Zn.  
However, based on the specific heat of the nonmagnetic reference compound LuCo$_2$Zn$_{20}$ (Fig.~\ref{fig:C_LCZ}), the contributions from Co and Zn, $C^\mathrm{nuc}_\mathrm{Co} + C^\mathrm{nuc}_\mathrm{Zn}$, are clearly negligible compared with that from the Ho site.  
Details of the LuCo$_2$Zn$_{20}$ measurements are provided in Appendix~\ref{app_subsec:C_LuCo2Zn20_results}. 
To isolate the contribution of the 4$f$ electrons and Ho nuclear spins, denoted by $C_\mathrm{Ho}$, we subtracted the electronic and phonon contribution of LuCo$_2$Zn$_{20}$, $C_\mathrm{nonmag}$, together with $C^\mathrm{nuc}_\mathrm{Co} + C^\mathrm{nuc}_\mathrm{Zn}$, from the total specific heat $C$:  
\begin{equation}
C_\mathrm{Ho} = C - (C_\mathrm{nonmag} + C^\mathrm{nuc}_\mathrm{Co} + C^\mathrm{nuc}_\mathrm{Zn}).
\label{eq:HCZ_C_component}
\end{equation}
The resulting $C_\mathrm{Ho}(T)$ at 0, 1, 4, and 9~T for $H \parallel [110]$ is shown in Fig.~\ref{fig:C_comparison}(a) and is used for the level-scheme analysis in Sec.~\ref{sec:analysis}.  
The procedure for estimating $C_\mathrm{nonmag} + C^\mathrm{nuc}_\mathrm{Co} + C^\mathrm{nuc}_\mathrm{Zn}$ from LuCo$_2$Zn$_{20}$ is described in Appendix~\ref{app_subsec:C_bg}.

%%%%%%%%%%%%%%%%%%%%%%%%%%%%%%%%%%%%%%%
%%%  Table 2
%%%%%%%%%%%%%%%%%%%%%%%%%%%%%%%%%%%%%%%
\begin{table}[tb!]
\caption{\label{tab:list_nuclear_spin}
Isotopes with nuclear spins for Ho, Lu, Co, and Zn.  
Here, $n_a$, $I$, and $g_\mathrm{N}$ denote the natural abundance, nuclear spin, and nuclear $g$ factor, respectively.  
Values of $n_a$ and $I$ are taken from Ref.~\cite{Rumble2024}, while $g_\mathrm{N}$ is calculated as $g_\mathrm{N} = \mu_n/I$, where $\mu_n$ is the nuclear magnetic moment given in units of the nuclear magneton $\mu_\mathrm{N}$ in Ref.~\cite{Rumble2024}.}
\catcode`!=\active \def!{\phantom{.}}
\catcode`?=\active \def?{\phantom{0}}
\begin{ruledtabular}
\begin{tabular}{lccl}
Isotope &
$n_a$ &
$I$ &
\multicolumn{1}{c}{$g_\mathrm{N}$} \\
\midrule
$^{165}$Ho & 1 & 7/2 & 1.192 \\
$^{175}$Lu & 0.974 01 & 7/2 & 0.637 91 \\
$^{176}$Lu & 0.025 99 & 7 & 0.4527 \\
$^{59}$Co & 1 & 7/2 & 1.322 \\
$^{67}$Zn & 0.0404 ? & 5/2 & 0.350 082 \\
\end{tabular}
\end{ruledtabular}
\end{table}
%%%%%%%%%%%%%%%%%%%%%%%%%%%%%%%%%%%%%%%
%%%
%%%%%%%%%%%%%%%%%%%%%%%%%%%%%%%%%%%%%%%

\subsection{\label{subsec:phase_transition} Phase transition of \boldmath HoCo$_2$Zn$_{20}$}

\begin{figure}[tb!]
\includegraphics[keepaspectratio, width=8.5cm, clip]{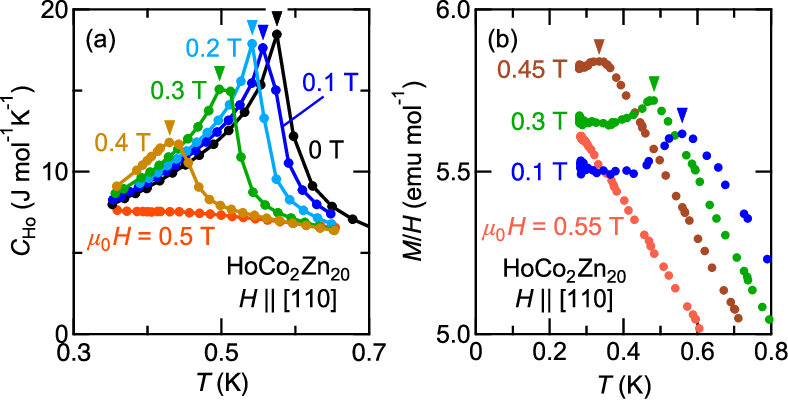}
\caption{\label{fig:phase_transition} 
Phase transition in HoCo$_2$Zn$_{20}$ observed in (a) the specific heat contribution from Ho sites, $C_\mathrm{Ho}$, and (b) the magnetization divided by the applied magnetic field, $M/H$. 
The transition temperatures, indicated by downward-pointing triangles, are defined as the temperatures at which $C_\mathrm{Ho}(T)$ or $M(T)/H$ exhibits a maximum.
}
\end{figure}

\begin{figure}[tb!]
\includegraphics[keepaspectratio, width=7cm, clip]{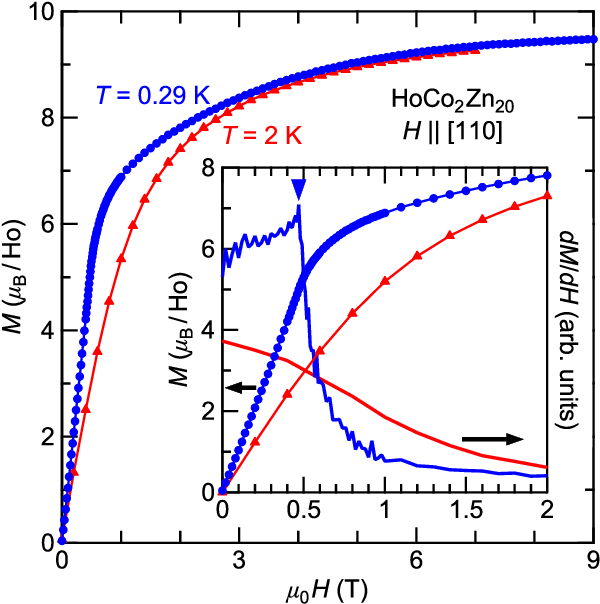}
\caption{\label{fig:MH_faraday}
Magnetization curves $M(H)$ of HoCo$_2$Zn$_{20}$ measured at 0.29 and 2~K for $H \parallel [110]$. 
The inset shows an enlarged view of $M(H)$ (lines with symbols, left axis) and the differential magnetization $dM/dH$ (lines without symbols, right axis).
}
\end{figure}

\begin{figure}[tb!]
\includegraphics[keepaspectratio, width=6cm, clip]{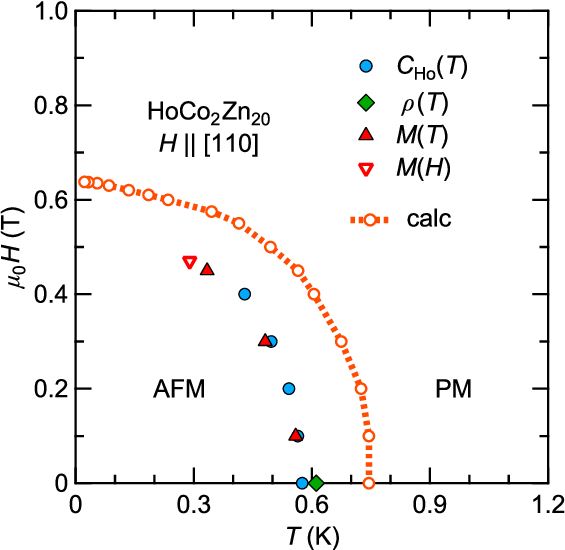}
\caption{\label{fig:phase_diagram} $H$--$T$ phase diagram of HoCo$_2$Zn$_{20}$ for $H \parallel [110]$.  
AFM and PM denote the antiferromagnetic and paramagnetic states, respectively.  
Open circles indicate transition temperatures calculated from Eq.~\eqref{eq:Hamil_MF_total} using the refined parameters given in Secs.~\ref{subsec:cef_refinement} and \ref{subsec:hyperfine_refinement}.}
\end{figure}

To investigate the nature of the low-temperature phase below 0.6~K, we performed specific heat and magnetization measurements under weak magnetic fields.
Figure~\ref{fig:phase_transition}(a) shows $C_\mathrm{Ho}(T)$ for $H \parallel [110]$ in steps of 0.1~T.  
The sharp peak observed at zero field shifts to lower temperatures and broadens as the magnetic field increases.  
At 0.5~T, the peak in $C_\mathrm{Ho}(T)$ is no longer observed down to the base temperature of 0.35~K.
In the temperature dependence of the magnetization $M(T)$ for $H \parallel [110]$ at 0.1 and 0.3~T [Fig.~\ref{fig:phase_transition}(b)], a broad maximum appears at the same temperature where $C_\mathrm{Ho}(T)$ shows a peak.  
This maximum in $M(T)$, also observed at 0.45~T, suggests the onset of antiferromagnetic (AFM) order.  
At 0.55~T, $M(T)$ increases monotonically upon cooling down to the base temperature of 0.28~K.
%%%%%%%%%%%%%%%%%%%%%%%%%%%%%%%%%%%%%%%
%%%  Fig. 3
%%%%%%%%%%%%%%%%%%%%%%%%%%%%%%%%%%%%%%%
%%%%%%%%%%%%%%%%%%%%%%%%%%%%%%%%%%%%%%%
%%%
%%%%%%%%%%%%%%%%%%%%%%%%%%%%%%%%%%%%%%%
We also measured the field dependence of the magnetization $M(H)$ along the $[110]$ direction at 0.29 and 2~K, as presented in Fig.~\ref{fig:MH_faraday}, via the capacitive Faraday method~\cite{Sakakibara1994}. 
While the differential magnetization $dM/dH$ at 2~K decreases monotonically, $dM/dH$ at 0.29~K gradually increases and exhibits a peak near 0.47~T (inset of Fig.~\ref{fig:MH_faraday}), suggesting a weak metamagnetic transition from the AFM to the paramagnetic state.
%
%%%%%%%%%%%%%%%%%%%%%%%%%%%%%%%%%%%%%%%
%%%  Fig. 4
%%%%%%%%%%%%%%%%%%%%%%%%%%%%%%%%%%%%%%%
%%%%%%%%%%%%%%%%%%%%%%%%%%%%%%%%%%%%%%%
%%%
%%%%%%%%%%%%%%%%%%%%%%%%%%%%%%%%%%%%%%%
The magnetic field-temperature ($H–T$) phase diagram for $H \parallel [ 110 ]$, constructed from $\rho(T)$, $C_\mathrm{Ho}(T)$, $M(T)$, and $M(H)$ data, is shown in Fig.~\ref{fig:phase_diagram}.
%
%%%%%%%%%%%%%%%%%%%%%%%%%%%%%%%%%%%%%%%
%%%  Fig. 5
%%%%%%%%%%%%%%%%%%%%%%%%%%%%%%%%%%%%%%%
%%%%%%%%%%%%%%%%%%%%%%%%%%%%%%%%%%%%%%%
%%%
%%%%%%%%%%%%%%%%%%%%%%%%%%%%%%%%%%%%%%%

Based on Figs.~\ref{fig:chiT}--\ref{fig:phase_diagram} and Table~\ref{tab:CW_fit_results}, we can conclude that the low-temperature phase of HoCo$_2$Zn$_{20}$ below 0.6~K at zero field is an AFM phase for the following two main reasons.  
First, $M(T)$ exhibits features consistent with AFM behavior across the entire measured temperature range.  
The Curie-Weiss temperature $\theta_\mathrm{C}$ is negative (Table~\ref{tab:CW_fit_results}), and $M(T)$ decreases with decreasing temperature below the transition temperature [Fig.~\ref{fig:phase_transition}(b)].  
In addition, as shown in the inset of Fig.~\ref{fig:chiT}, $\chi(T)$ below approximately 10~K is slightly suppressed compared to the Curie-Weiss fit performed between 100 and 300~K, suggesting the development of antiferromagnetic correlations among Ho moments upon cooling.  
Second, the basic physical properties of HoCo$_2$Zn$_{20}$ reported in this section (Sec.~\ref{sec:results}) closely resemble those of the antiferromagnet NdCo$_2$Zn$_{20}$.  
NdCo$_2$Zn$_{20}$, an isostructural compound of HoCo$_2$Zn$_{20}$, exhibits AFM order with a N\'{e}el temperature of $T_\mathrm{N} = 0.53$~K~\cite{Yamamoto2019}, as confirmed microscopically by neutron scattering experiments~\cite{Yamamoto2023}.  
In fact, the magnetic specific heat $C_\mathrm{m}(T)$, resistivity $\rho(T)$, and magnetic susceptibility $\chi(T)$ near $T_\mathrm{N}$ in NdCo$_2$Zn$_{20}$ [see Figs.~5(a), 6(a), and 7 in Ref.~\cite{Yamamoto2019}, respectively] exhibit qualitatively similar behavior to $C_\mathrm{Ho}(T)$, $\rho(T)$, and $M(T)/H$ in HoCo$_2$Zn$_{20}$.  
Moreover, the weak metamagnetic transition (inset of Fig.~\ref{fig:MH_faraday}) and the $H$--$T$ phase diagram (Fig.~\ref{fig:phase_diagram}) observed in HoCo$_2$Zn$_{20}$ closely resemble those in NdCo$_2$Zn$_{20}$ [see the inset of Fig.~7 and Fig.~5(a) in Ref.~\cite{Yamamoto2019}, respectively]. 
Taken together, the phase transition of HoCo$_2$Zn$_{20}$ is understood as an AFM transition of the Ho magnetic moments.

%
%
%
%%%%%%%%%%%%%%%%%%%%%%%%%%%%%%%%%%%%%%%
\section{\label{sec:analysis}Analysis of the local electron-nuclear state}
%%%%%%%%%%%%%%%%%%%%%%%%%%%%%%%%%%%%%%%

In this section, we determine the electron-nuclear level scheme at the Ho sites based on the experimental results of magnetization and specific heat.

\subsection{Model}

Based on recent studies of other Ho-based magnetic materials~\cite{Wendl2022, Gronemann2023}, we construct the following Hamiltonian to calculate thermodynamic quantities and determine the electron-nuclear level scheme at the Ho sites:
\begin{equation}
\begin{aligned}
    \mathcal{H} =& 
        \sum_{i} \mathcal{H}_\mathrm{CEF}^\mathrm{cubic}
        + \mu_0 \mu_\mathrm{B} g_J \bm{H} \cdot \sum_{i} \bm{J}_i
        + \mu_0 \mu_\mathrm{N} g_\mathrm{N} \bm{H} \cdot \sum_{i} \bm{I}_i \\
        & + A_\mathrm{HF} \sum_{i} \bm{I}_i \cdot \bm{J}_i
        - \mathcal{J} \sum_{\langle i,j \rangle} \bm{J}_i \cdot \bm{J}_j.
\end{aligned}
\label{eq:Hamil_HCZ}
\end{equation}
Here, the indices $i$ label Ho sites and $\langle i, j \rangle$ denotes nearest-neighbor pairs.  
The five terms on the right-hand side of Eq.~\eqref{eq:Hamil_HCZ} correspond, respectively, to the crystalline electric field (CEF) term, the Zeeman term for $4f$ electrons, the nuclear Zeeman term, the hyperfine interaction, and the intersite exchange interaction.  
Details of each term are given below.

The term $\mathcal{H}_\mathrm{CEF}^\mathrm{cubic}$ represents the CEF Hamiltonian for the Ho$^{3+}$ ion in cubic symmetry and is expressed as~\cite{Lea1962}
\begin{equation}
    \mathcal{H}_\mathrm{CEF}^\mathrm{cubic} = 
    W \left\{ x \frac{O_4^0 + 5 O_4^4}{60} + \left( 1 - |x| \right) \frac{O_6^0 - 21 O_6^4}{13 \ 860} \right\},
\end{equation}
where $W$ and $x$ are CEF parameters, and $O_n^m$ are Stevens operators~\cite{Stevens1952, Hutchings1964}.
The second and third terms describe the Zeeman interactions of the $4f$ electrons and the $^{165}$Ho nuclear spins, respectively.  
Here, $\mu_\mathrm{B}$ is the Bohr magneton, $g_J = 5/4$ is the Land\'{e} $g$ factor for the $4f^{10}$ configuration, $\mu_\mathrm{N}$ is the nuclear magneton, and $g_\mathrm{N} = 1.192$ is the nuclear $g$ factor for $^{165}$Ho (Table \ref{tab:list_nuclear_spin}).
The fourth term represents the hyperfine interaction between the electronic angular momentum $\bm{J}$ and the nuclear spin $\bm{I}$ at each Ho site, with $A_\mathrm{HF}$ denoting the hyperfine coupling constant. 
As noted in Sec.~\ref{sec:introduction}, nuclear electric quadrupole interactions are neglected due to the cubic symmetry of the Ho site~\cite{Abragam1961}.
Finally, the fifth term describes an isotropic Heisenberg exchange interaction between nearest-neighbor Ho magnetic moments with the exchange constant $\mathcal{J}$.

By applying the mean-field approximation  
$\bm{J}_i \cdot \bm{J}_j = \bm{J}_i \cdot \langle \bm{J}_j \rangle + \langle \bm{J}_i \rangle \cdot \bm{J}_j - \langle \bm{J}_i \rangle \cdot \langle \bm{J}_j \rangle$,  
where $\langle \bm{J}_i \rangle$ denotes the canonical thermal average of the operator $\bm{J}_i$,  
the total Hamiltonian $\mathcal{H}$ for $\mathcal{J} < 0$ can be written as
\begin{gather}
    \mathcal{H} = \sum_{i \in A} \mathcal{H}_A + \sum_{i \in B} \mathcal{H}_B,
    \label{eq:Hamil_MF_total} \\
    \begin{aligned}
        \mathcal{H}_\alpha = & 
        \mathcal{H}_\mathrm{CEF}^\mathrm{cubic}
        + \mu_0 \left( g_J \mu_\mathrm{B} \bm{J}_\alpha 
        + g_\mathrm{N} \mu_\mathrm{N} \bm{I}_\alpha \right) \cdot \bm{H} \\
        & + A_\mathrm{HF} \bm{I}_\alpha \cdot \bm{J}_\alpha
        - J_\mathrm{ex} \bm{J}_\alpha \cdot \langle \bm{J}_\beta \rangle
        + \frac{J_\mathrm{ex}}{2} \langle \bm{J}_\alpha \rangle \cdot \langle \bm{J}_\beta \rangle,
    \end{aligned}
    \label{eq:Hamil_MF_single}
\end{gather}
where $A$ and $B$ denote the two sublattices of the Ho sites, $(\alpha, \beta)$ refers to either $(A, B)$ or $(B, A)$, and $J_\mathrm{ex} = z \mathcal{J}$, with $z$ being the number of nearest-neighbor sites. 
Since the Ho sites in HoCo$_2$Zn$_{20}$ form a diamond lattice, we set $z = 4$. 
In the case of $\mathcal{J} > 0$ ($J_\mathrm{ex} > 0$), the term $\sum_{i \in B} \mathcal{H}_B$ in Eq.~\eqref{eq:Hamil_MF_total} is omitted because the system consists of a single sublattice. 
Accordingly, $(\alpha, \beta)$ in Eq.~\eqref{eq:Hamil_MF_single} becomes $(A, A)$. 
In this section, we employ the mean-field Hamiltonian given by Eqs.~\eqref{eq:Hamil_MF_total} and \eqref{eq:Hamil_MF_single}, using four free parameters, $W$, $x$, $J_\mathrm{ex}$, and $A_\mathrm{HF}$, to refine the quantum states at the Ho sites.

\subsection{\label{subsec:cef_refinement} Refinement of CEF parameters and the magnetic exchange constant}

\begin{figure}[tb!]
\includegraphics[keepaspectratio, width=8cm, clip]{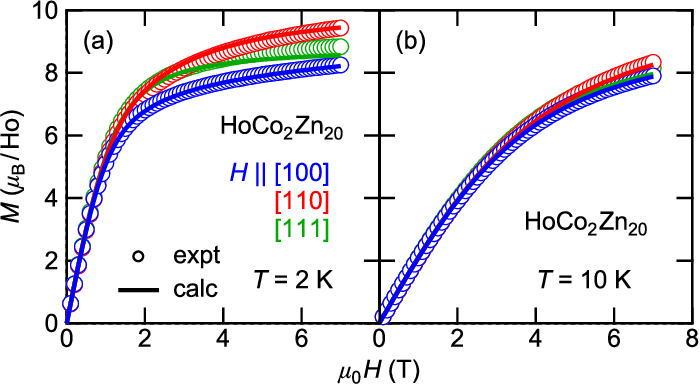}
\caption{\label{fig:MH_2K10K} Magnetization curves of HoCo$_2$Zn$_{20}$ at (a) 2~K and (b) 10~K. Open circles represent the experimental data, and solid lines indicate the calculated results based on the Hamiltonian excluding nuclear spin terms (see Sec.~\ref{subsec:cef_refinement} for details).}
\end{figure}

\begin{figure}[tb!]
\includegraphics[keepaspectratio, width=7.5cm, clip]{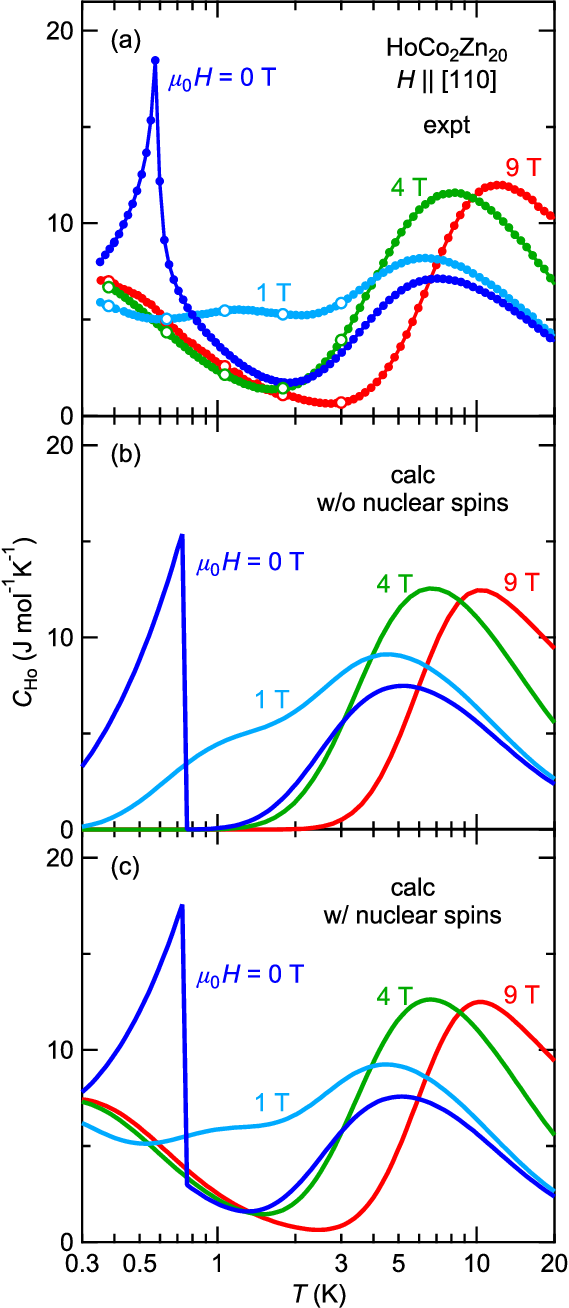}
\caption{\label{fig:C_comparison}
Temperature dependence of the specific heat contribution from Ho ions, $C_\mathrm{Ho}$, in HoCo$_2$Zn$_{20}$.  
Panels (a), (b), and (c) show the experimental data, the calculated results without nuclear spins, and the calculated results including nuclear spins, respectively.  
Open circles in panel (a) indicate the data points used to refine the hyperfine coupling constant $A_\mathrm{HF}$ (see Appendix~\ref{app_subsec:analysis_method_2} for details).}
\end{figure}

We first determine the CEF parameters $W$ and $x$, as well as the magnetic exchange parameter $J_\mathrm{ex}$, based on the isothermal magnetization data at 2 and 10~K shown in Fig.~\ref{fig:MH_2K10K}.  
Since the enhancement of $C_\mathrm{Ho}$ due to nuclear spin contributions becomes significant below approximately 2~K [Fig.~\ref{fig:C_comparison}(a)], the Hamiltonian excluding nuclear spin terms is expected to be valid above this temperature.  
Therefore, $W$, $x$, and $J_\mathrm{ex}$ were determined using the total Hamiltonian $\mathcal{H}$ with the nuclear Zeeman term $\mu_0 g_\mathrm{N} \mu_\mathrm{N} \bm{I}_\alpha \cdot \bm{H}$ and the hyperfine interaction term $A_\mathrm{HF} \bm{I}_\alpha \cdot \bm{J}_\alpha$ omitted from Eq.~\eqref{eq:Hamil_MF_single}.  
Details of the analysis method are provided in Appendix~\ref{app_subsec:analysis_method_1}.  
The best-fit parameters obtained from the analysis of $M(H)$ at 2 and 10~K are $W = 0.0443$~K, $x = -0.0640$, and $J_\mathrm{ex} = -0.0511$~K.  
The corresponding calculated $M(H)$ curves are shown as solid lines in Fig.~\ref{fig:MH_2K10K}.  
Although $J_\mathrm{ex}$ was treated as a free parameter that could take either positive or negative values, a negative value was obtained, consistent with the antiferromagnetic nature of HoCo$_2$Zn$_{20}$.

%%%%%%%%%%%%%%%%%%%%%%%%%%%%%%%%%%%%%%%
%%%  Fig. 6
%%%%%%%%%%%%%%%%%%%%%%%%%%%%%%%%%%%%%%%
%%%%%%%%%%%%%%%%%%%%%%%%%%%%%%%%%%%%%%%
%%%
%%%%%%%%%%%%%%%%%%%%%%%%%%%%%%%%%%%%%%%

%%%%%%%%%%%%%%%%%%%%%%%%%%%%%%%%%%%%%%%
%%%  Fig. 7
%%%%%%%%%%%%%%%%%%%%%%%%%%%%%%%%%%%%%%%
%%%%%%%%%%%%%%%%%%%%%%%%%%%%%%%%%%%%%%%
%%%
%%%%%%%%%%%%%%%%%%%%%%%%%%%%%%%%%%%%%%%

To evaluate the validity of these parameters, we calculated $C_\mathrm{Ho}(T)$ using the values of $W$, $x$, and $J_\mathrm{ex}$ determined above, as shown in Fig.~\ref{fig:C_comparison}(b).  
Although these parameters were optimized using the experimental data at 2 and 10~K, where HoCo$_2$Zn$_{20}$ is in the paramagnetic state, the calculated $C_\mathrm{Ho}(T)$ exhibits an antiferromagnetic transition at nearly the same temperature as observed experimentally.  
In addition, the Curie-Weiss temperature $\theta_\mathrm{C}$ estimated from the mean-field exchange constant $J_\mathrm{ex}$ via $\theta_\mathrm{C} = J(J+1)J_\mathrm{ex}/3$ is $-1.23$~K, which is close to the experimental value obtained from $\chi(T)$ (Table~\ref{tab:CW_fit_results}).  
The agreement between the calculated and experimental values of $T_\mathrm{N}$ and $\theta_\mathrm{C}$ supports the validity of $J_\mathrm{ex}$.  
Furthermore, the calculated $C_\mathrm{Ho}(T)$ in Fig.~\ref{fig:C_comparison}(b) semiquantitatively reproduces the experimental data in Fig.~\ref{fig:C_comparison}(a) above approximately 2~K, both at zero field and under magnetic fields, confirming the reliability of the CEF parameters.  
The discrepancy in $C_\mathrm{Ho}$ below 2~K can be attributed to the neglect of nuclear spin effects.  
Therefore, in the next step, we include the nuclear Zeeman and hyperfine interaction terms, $\mu_0 g_\mathrm{N} \mu_\mathrm{N} \bm{I}_\alpha \cdot \bm{H}$ and $A_\mathrm{HF} \bm{I}_\alpha \cdot \bm{J}_\alpha$, and refine the hyperfine coupling constant $A_\mathrm{HF}$ to reproduce the experimental $C_\mathrm{Ho}(T)$ down to the lowest temperature of 0.35~K.

\subsection{\label{subsec:hyperfine_refinement} Refinement of the hyperfine coupling constant}

$A_\mathrm{HF}$ was refined using the full Hamiltonian $\mathcal{H}$ in Eq.~\eqref{eq:Hamil_MF_total}, including all terms in Eq.~\eqref{eq:Hamil_MF_single}.  
In general, the specific heat calculated within the mean-field approximation exhibits a discontinuous jump at the transition temperature, as seen in $C_\mathrm{Ho}(T)$ at 0~T in Fig.~\ref{fig:C_comparison}(b), due to the neglect of magnetic fluctuations.  
In contrast, the experimentally observed $C_\mathrm{Ho}(T)$ at 0~T varies continuously, indicating the presence of magnetic fluctuations.  
Therefore, while the 0~T data were excluded from the refinement, $C_\mathrm{Ho}(T)$ data below 3~K at 1, 4, and 9~T were used to refine $A_\mathrm{HF}$. 
Since these data at finite magnetic fields are well separated from the phase boundary in the experimental $H$--$T$ phase diagram (Fig.~\ref{fig:phase_diagram}), magnetic fluctuations are expected to be negligible.
Therefore, the mean-field treatment is sufficient for estimating $A_\mathrm{HF}$.
Details of the analysis method are provided in Appendix~\ref{app_subsec:analysis_method_2}.
As a result of the refinement, the hyperfine coupling constant was determined to be $A_\mathrm{HF} = 0.0355$~K~\cite{AHF_note}.  
This value is close to the expected value of $A_\mathrm{HF} = 0.0390$~K for an isolated Ho$^{3+}$ ion~\cite{Bleaney1972}.  
The calculated $C_\mathrm{Ho}(T)$ using all refined parameters, $W$, $x$, $J_\mathrm{ex}$, and $A_\mathrm{HF}$, is shown in Fig.~\ref{fig:C_comparison}(c), and it reproduces the experimental $C_\mathrm{Ho}(T)$ well over the entire temperature range.  
We also note that not only the specific heat but also the magnetization (Fig.~\ref{fig:chiT_and_MH_calc}) and the $H$--$T$ phase diagram (Fig.~\ref{fig:phase_diagram}) are well reproduced using the refined parameters.

\subsection{\label{subsec:level_scheme} Level scheme of Ho sites at 0~T}

\begin{figure}[tb!]
\includegraphics[keepaspectratio, width=8.5cm, clip]{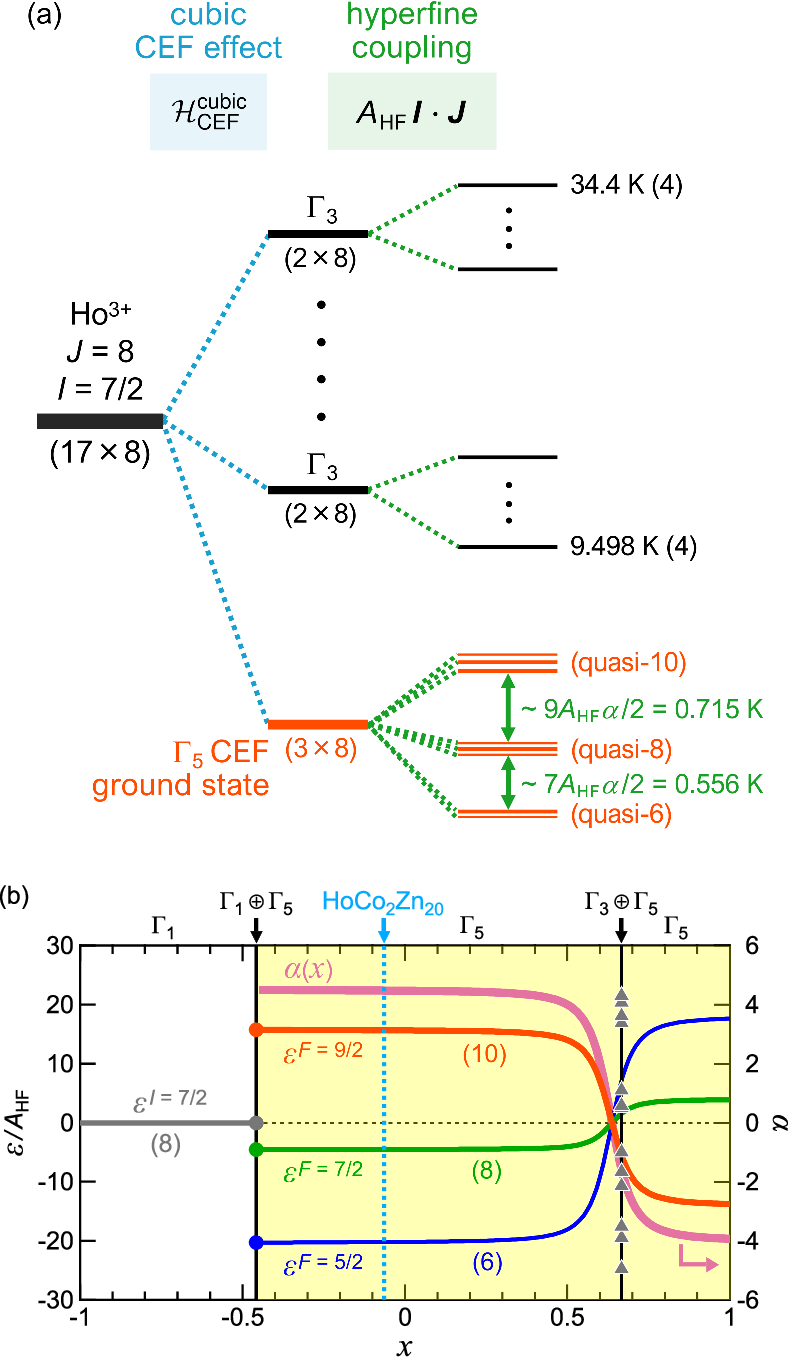}
\caption{\label{fig:HCZ_level_scheme} 
(a) Schematic energy-level diagram of the Ho sites in HoCo$_2$Zn$_{20}$. 
Numbers in parentheses indicate the degeneracy of each level; the prefix “quasi” denotes quasidegenerate levels. 
(b) Eigenenergy diagram as a function of $x$ for $W > 0$, restricted to the CEF ground-state subspace. 
Eigenenergies $\varepsilon$ (left axis) are normalized by the hyperfine coupling constant $A_\mathrm{HF}$. 
Degeneracies are again indicated in parentheses. 
The light-yellow background indicates the region where the CEF ground state is $\Gamma_5$, within which $\alpha(x)$ (light-purple line, right axis) is defined. 
The triangle at $x = 2/3$ denotes the eigenenergy in the direct-sum space $\Gamma_3 \oplus \Gamma_5$, where the degeneracy is 2 or 4 (see Appendix~\ref{app_sec:splitting_CEF}). 
The light-blue vertical line indicates the value of $x$ determined for HoCo$_2$Zn$_{20}$ ($x = -0.0640$).
}
\end{figure}

Since all free parameters in Eq.~\eqref{eq:Hamil_MF_total} have been determined, we calculate the energy-level scheme of the Ho sites at 0~T in the paramagnetic state by diagonalizing the single-site Hamiltonian at 0~T, given in Eq.~\eqref{eq:Hamil_MF_single} as $\mathcal{H}^\mathrm{0T} = \mathcal{H}_\mathrm{CEF}^\mathrm{cubic} + A_\mathrm{HF} \bm{I} \cdot \bm{J}$.
The resulting level scheme is shown in Fig.~\ref{fig:HCZ_level_scheme}(a).
In the absence of both CEF effects and hyperfine coupling, the ground state of the $4f^{10}$ configuration is a $J = 8$ multiplet, and the $^{165}$Ho nuclear spin $I = 7/2$ remains degenerate.
Thus, the ground state consists of $17 \times 8 = 136$ degenerate eigenstates.

%%%%%%%%%%%%%%%%%%%%%%%%%%%%%%%%%%%%%%%
%%%  Fig. 8
%%%%%%%%%%%%%%%%%%%%%%%%%%%%%%%%%%%%%%%
%%%%%%%%%%%%%%%%%%%%%%%%%%%%%%%%%%%%%%%
%%%
%%%%%%%%%%%%%%%%%%%%%%%%%%%%%%%%%%%%%%%

Under the cubic CEF, this multiplet in the $J$ space splits into a $\Gamma_1$ singlet, two $\Gamma_3$ doublets, two $\Gamma_4$ triplets, and two $\Gamma_5$ triplets \cite{Lea1962}.
In HoCo$_2$Zn$_{20}$, the CEF ground state corresponds to the irreducible representation $\Gamma_5$. 
Since the CEF Hamiltonian does not include the nuclear spin operator $\bm{I}$, nuclear spin degeneracy is preserved, resulting in $3 \times 8 = 24$ degenerate eigenstates in the CEF ground state.
It is noteworthy that the energy separation between the $\Gamma_5$ ground state and the highest CEF level is only 33.1~K. 
Such a narrow level spacing is a characteristic feature of the CeCr$_2$Al$_{20}$-type structure, where the rare-earth ion is enclosed in an almost spherical cage formed by 16 Zn atoms \cite{Nasch1997}.
Indeed, the CEF splitting $\Delta_\mathrm{CEF}$ confirmed by inelastic neutron scattering experiments is approximately 80~K in NdCo$_2$Zn$_{20}$ \cite{Yamamoto2023a} and 30~K in YbCo$_2$Zn$_{20}$ \cite{Kaneko2012}.

The 24-fold degenerate CEF ground state is further split by the hyperfine coupling into four quartets and four doublets.  
The corresponding eigenenergies are 0~K (2), 0.83~mK (4), 0.543~K (2), 0.546~K (4), 0.551~K (2), 1.256~K (4), 1.263~K (4), and 1.265~K (2), where the numbers in parentheses denote degeneracy.
One can identify quasisextet, quasioctet, and quasidectet structures near 0, 0.55, and 1.26~K, respectively. 
We use the prefix ``quasi'' to emphasize that these structures are not perfectly degenerate multiplets but are slightly split into doublets and quartets on the millikelvin or submillikelvin energy scale.
The resulting doublet and quartet states are eigenstates of $\mathcal{H}^\mathrm{0T}$.
The origin of these quasimultiplets can be understood from the hyperfine coupling within the $\Gamma_5$ CEF ground-state subspace, as explained below. 
We note that a similar situation was discussed for a cubic Pr compound in Ref.~\cite{Aoki2011}.

Since the total hyperfine splitting within the $\Gamma_5$ ground state (1.265~K) is much smaller than the energy gap between the ground pseudosextet and the lowest multiplet split from the first excited CEF level [9.498~K; see Fig.~\ref{fig:HCZ_level_scheme}(a)], it is justified to restrict the discussion to the $\Gamma_5$ subspace.
As detailed in Appendix \ref{app_sec:splitting_CEF}, the angular momentum operator in this subspace can be expressed as $\bm{J} = \alpha(x) \bm{S}$, where $\alpha(x)$ depends on the cubic CEF parameter $x$, and $\bm{S}$ is an effective spin operator with $S = 1$.
Accordingly, the hyperfine coupling becomes $A_\mathrm{HF} \bm{I} \cdot \bm{J} = A_\mathrm{HF} \alpha(x) \bm{I} \cdot \bm{S}$.
By defining the total angular momentum $\bm{F} = \bm{I} + \bm{S}$, the 24-fold degenerate ground state is split into multiplets labeled by $F = 7/2 - 1$, $7/2$, and $7/2 + 1$, corresponding to a sextet ($F = 5/2$), an octet ($F = 7/2$), and a dectet ($F = 9/2$), respectively.
Since \( \bm{I} \cdot \bm{S} = \big(\bm{F}^2 - \bm{I}^2 - \bm{S}^2\big)/2 \), the eigenenergy for each $F$ multiplet is given by
\begin{equation}
    \begin{aligned}
        A_\mathrm{HF} \alpha(x) \frac{F(F+1) - I(I+1) - S(S+1)}{2} \\
        = \frac{A_\mathrm{HF} \alpha(x)}{2} \left\{F(F+1) - \frac{33}{2}\right\}.
    \end{aligned}
    \label{eq:HF_eigenenergy}
\end{equation}

For the $^{165}$Ho isotope, $A_\mathrm{HF}$ is positive because, in the $f^{10}$ configuration, it is given by $A_\mathrm{HF} = 23 g_\mathrm{N} \mu_\mathrm{N} \mu_\mathrm{B} \braket{r^{-3}}/15$ \cite{Wybourne_note, Wybourne1965}, where $ g_\mathrm{N} = 1.192$ (Table \ref{tab:list_nuclear_spin}) and $\braket{r^{-3}}$ is the average inverse cubic radius of the $4f$ orbital.
Therefore, for $\alpha(x) > 0$, the ground state is the $F = 5/2$ sextet, whereas for $\alpha(x) < 0$, it becomes the $F = 9/2$ dectet.
In HoCo$_2$Zn$_{20}$, where $\alpha(x = -0.0640) = 4.479$, the ground state is the $F = 5/2$ sextet. 
The first and second excited states are the $F = 7/2$ octet and $F = 9/2$ dectet, located at $7A_\mathrm{HF} \alpha/2 = 0.556$~K and $8A_\mathrm{HF} \alpha = 1.271$~K above the ground state, respectively.
These eigenenergies closely match the quasioctet ($\sim 0.55$~K) and quasidectet levels ($\sim 1.26$~K) obtained from the numerical diagonalization of $\mathcal{H}^\mathrm{0T}$.
The further splitting of each $F$ multiplet into doublets and quartets on the millikelvin or submillikelvin scale likely originates from a small but finite occupancy of excited CEF states by $f$ electrons.

%
%
%
%%%%%%%%%%%%%%%%%%%%%%%%%%%%%%%%%%%%%%%
\section{\label{sec:discussion}Discussion}
%%%%%%%%%%%%%%%%%%%%%%%%%%%%%%%%%%%%%%%

\subsection{\label{subsec:eigenenergy_diagram} Eigenenergy diagram for \boldmath $W > 0$}

To investigate the hyperfine level scheme in general cubic Ho compounds, we calculated $\alpha(x)$ and constructed the eigenenergy diagram over a wide range of $x$. 
Figure~\ref{fig:HCZ_level_scheme}(b) shows the $x$ dependence of the eigenenergies of $A_\mathrm{HF} \bm{I} \cdot \bm{J}$ normalized by $A_\mathrm{HF}$, i.e., $\varepsilon/A_\mathrm{HF}$, for $W > 0$ within the CEF ground-state subspace, together with the variation of $\alpha(x)$.  
Details of the construction of this figure are provided in Appendix~\ref{app_sec:splitting_CEF}.  
The $\Gamma_5$ state is the ground level for $-38/83 < x \leq 1$, except at $x = 2/3$.  
Within this range, $\alpha(x)$ changes sign at $x = x_0 \ (\sim 0.64)$, where $\alpha(x_0) = 0$, and the $F = 9/2$ dectet becomes the ground state for $x > x_0$.  
This result demonstrates that the hyperfine level scheme is governed not only by $A_\mathrm{HF}$, which sets the overall energy scale of the hyperfine splitting, but also by the CEF parameter $x$.

\subsection{\label{subsec:MCK} Multichannel Kondo effect in materials containing nuclear spins}

Finally, we comment on the multichannel Kondo (MCK) effect in realistic materials.  
The MCK effect, in which a local moment at a magnetic site is overscreened by electrons from multiple conduction bands~\cite{Nozieres1980}, has attracted considerable attention because of the associated residual entropy~\cite{Tsvelick1985, Desgranges1985, Sacramento1991}, which reflects the presence of exotic quasiparticles such as Majorana and Fibonacci anyons~\cite{Han2022}.  
In bulk systems, experimental evidence for the MCK effect has recently been reported in the diluted Pr compound Y$_{1-x}$Pr$_x$Ir$_2$Zn$_{20}$ ($x \ll 1$)~\cite{Yamane2018a}, based on ultrasonic~\cite{Yanagisawa2019}, thermal expansion~\cite{Worl2022}, and magnetization~\cite{Yamane2025} measurements in the temperature range of $10^1-10^2$~mK.  
In this compound, the MCK effect is interpreted as a quadrupole Kondo effect, in which the electric quadrupole moment of the localized $f$ electrons is overscreened by conduction electrons~\cite{Cox1986}.  
The CEF ground state of the Pr ions in Y$_{1-x}$Pr$_x$Ir$_2$Zn$_{20}$ is a nonmagnetic $\Gamma_3$ doublet~\cite{Yamane2018a, Yanagisawa2019, Yamane2025}, which carries an active electric quadrupole moment but no magnetic dipole moment.  
Furthermore, the first excited CEF state lies about 30~K above the $\Gamma_3$ ground state~\cite{Yamane2018a}, which justifies restricting the discussion to the $\Gamma_3$ subspace when considering hyperfine splitting of the CEF ground state.  
Consequently, although natural Pr has a nuclear spin of $I = 5/2$ with 100\% natural abundance~\cite{Rumble2024}, the $\Gamma_3$ ground state does not split under hyperfine coupling.  
Thus, the influence of hyperfine coupling can be safely neglected in the quadrupole Kondo effect of Y$_{1-x}$Pr$_x$Ir$_2$Zn$_{20}$. 

In contrast, the magnetic MCK effect, in which a localized magnetic dipole is overscreened, has been theoretically proposed in several systems~\cite{Cox1993, Hotta2017}, and more recently in cubic Ho-based compounds~\cite{Hotta2021, Hotta2025}.  
According to Ref.~\cite{Hotta2021}, the magnetic MCK effect can be realized in cubic Ho-based compounds when the CEF ground state is the $\Gamma_5$ triplet.  
Since HoCo$_2$Zn$_{20}$ has a $\Gamma_5$ CEF ground state, diluted Ho systems R${'}_{1-x}$Ho$_x$Co$_2$Zn$_{20}$ ($x \ll 1$), in which Ho is substituted by a nonmagnetic element R$'$ to suppress magnetic order, are also expected to retain the $\Gamma_5$ CEF ground state. 
However, due to the presence of hyperfine coupling at magnetic sites—which has not been considered in theoretical studies of the MCK effect—the $\Gamma_5$ CEF states are no longer exact eigenstates except under special conditions~\cite{HF_splitting_note}. 
Whether the magnetic MCK effect can occur in the presence of hyperfine coupling remains an open question, both experimentally and theoretically. 
In the case of cubic Ho compounds, the diluted system R${'}_{1-x}$Ho$_x$Co$_2$Zn$_{20}$ may serve as a promising candidate for testing the realization of the MCK effect in a system with an $F = 5/2$ sextet ground state. 
In Appendix~\ref{app_sec:residual_entropy}, we propose a detailed experimental procedure for capturing the residual entropy associated with the MCK effect in diluted cubic systems containing nuclear spins.
At the same time, theoretical investigations are required to explore the possibility of the MCK effect when the hyperfine-split ground state is either an $F = 5/2$ sextet or an $F = 9/2$ dectet. 

%
%
%
%%%%%%%%%%%%%%%%%%%%%%%%%%%%%%%%%%%%%%%
\section{\label{sec:conclusion}Conclusion}
%%%%%%%%%%%%%%%%%%%%%%%%%%%%%%%%%%%%%%%

In this study, we revealed that HoCo$_2$Zn$_{20}$ exhibits antiferromagnetic order and determined its CEF parameters, magnetic exchange constant, and hyperfine coupling constant using macroscopic thermodynamic probes.  
These refined parameters clarified the hyperfine level scheme in the paramagnetic state, where the ground state is identified as an $F = 5/2$ quasisextet formed by the coupling between the $f$ electron and the $^{165}$Ho nuclear spin.  
It is noteworthy that the energy width of the CEF ground-state splitting due to hyperfine coupling exceeds 1~K at 0~T. 
This finding indicates that, if an Ho-based compound remains in the paramagnetic state down to subkelvin temperatures at 0~T, understanding the electron-nuclear level scheme may be essential for interpreting phenomena at very low temperatures, such as magnetic MCK effects~\cite{Hotta2021, Hotta2025} and frustrated magnetism~\cite{Stockert2020, Boraley2025}.  
Furthermore, the eigenenergy diagram of the hyperfine coupling within the subspace of the CEF ground state shows that the CEF parameters can be critically important in determining the low-energy electron-nuclear level scheme. 
We believe that our study on HoCo$_2$Zn$_{20}$ represents a first step toward determining hyperfine electron-nuclear entangled states via macroscopic measurements in a variety of materials with nuclear spins.

\begin{acknowledgments}
We thank K.~Ota and S.~Ichioka for assistance with single-crystal x-ray diffraction, and T.~Hotta for fruitful discussions.  
This work was largely supported by JSPS Grants-in-Aid for Scientific Research (KAKENHI) Grants No.~JP23H04866, No.~JP23H04870, and No.~JP24K22867.  
Additional support was provided by the JST FOREST Program (No.~JPMJFR2233), JSPS KAKENHI No.~JP23H04871, No.~JP23K03330, No.~JP23K03332, No.~JP23K25829, No.~JP24K00574, and No.~JP24H01673, and by the Motizuki Fund of the Yukawa Memorial Foundation. 
\end{acknowledgments}

\appendix

%
%
%
%%%%%%%%%%%%%%%%%%%%%%%%%%%%%%%%%%%%%%%
\section{\label{app_sec:M_suppl} Supplemental figures of magnetization}
%%%%%%%%%%%%%%%%%%%%%%%%%%%%%%%%%%%%%%%

\begin{figure}[tb!]
\includegraphics[keepaspectratio, width=8.5cm, clip]{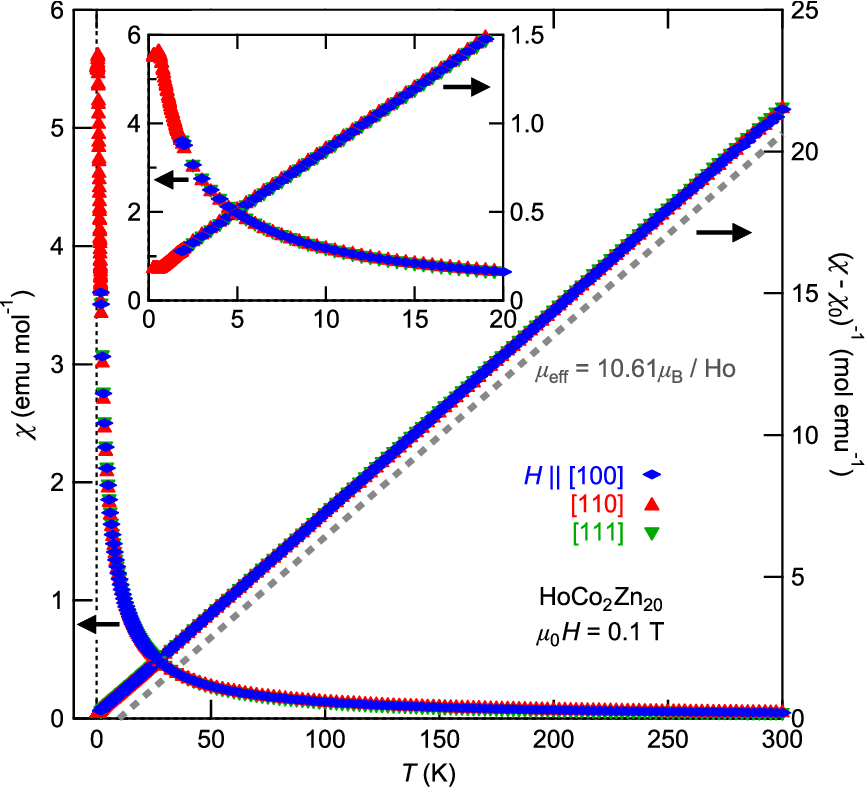}
\caption{\label{fig:chiT_all} 
Temperature dependence of $\chi$ (left axis) and $(\chi - \chi_0)^{-1}$ (right axis) in HoCo$_2$Zn$_{20}$ for $H \parallel [100]$, $[110]$, and $[111]$.  
The data for $H \parallel [110]$ are the same as those in Fig.~\ref{fig:chiT}.  
The gray dashed line indicates the expected slope of the inverse susceptibility for free Ho$^{3+}$ ions.  
The inset shows an enlarged view of $\chi$ and $(\chi - \chi_0)^{-1}$ below 20~K.}
\end{figure}

Figure~\ref{fig:chiT_all} shows the temperature dependence of $\chi$ and $(\chi - \chi_0)^{-1}$ along the [100], [110], and [111] directions, confirming the absence of magnetic anisotropy in weak magnetic fields.

%%%%%%%%%%%%%%%%%%%%%%%%%%%%%%%%%%%%%%%
%%%  Fig. 9
%%%%%%%%%%%%%%%%%%%%%%%%%%%%%%%%%%%%%%%
%%%%%%%%%%%%%%%%%%%%%%%%%%%%%%%%%%%%%%%

\begin{figure}[tb!]
\includegraphics[keepaspectratio, width=8.5cm, clip]{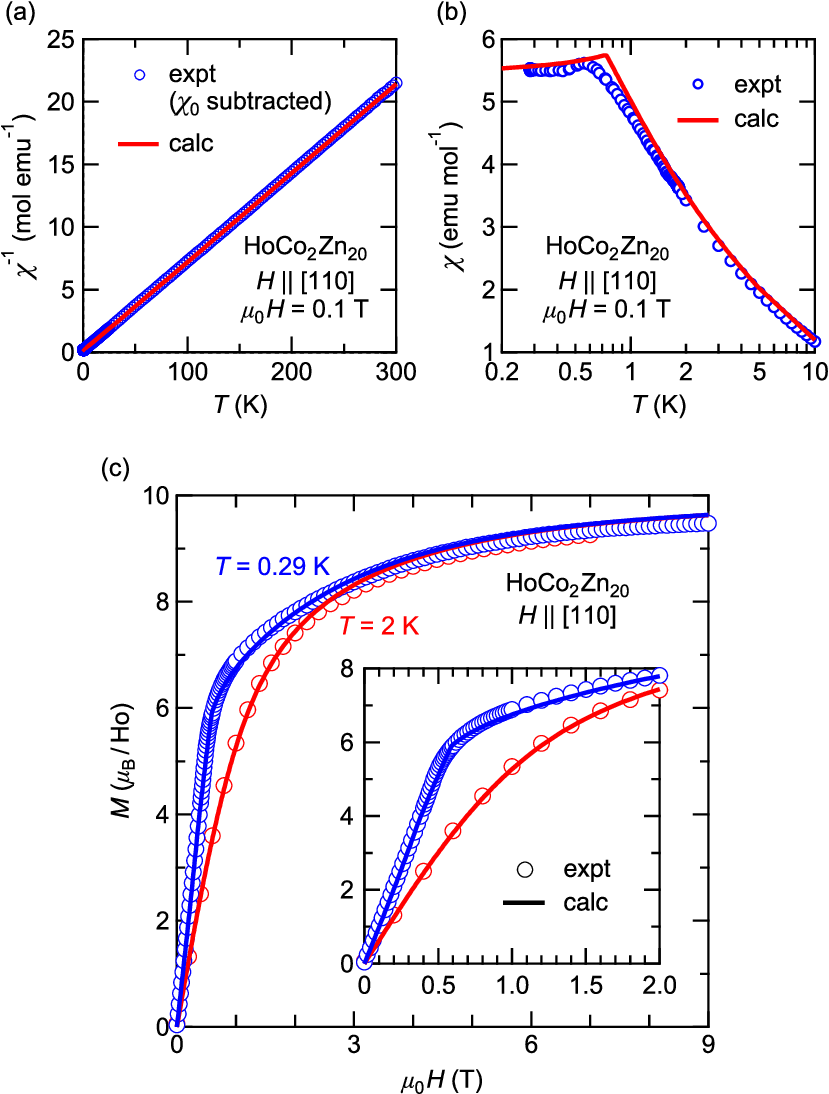}
\caption{\label{fig:chiT_and_MH_calc} 
Comparison of (a) $\chi^{-1}(T)$, (b) $\chi(T)$, and (c) $M(H)$ for $H \parallel [110]$ between experimental data (open circles) and calculations (solid lines) including all terms in Eq.~\eqref{eq:Hamil_MF_single}.  
The refined parameters ($W$, $x$, $J_\mathrm{ex}$, and $A_\mathrm{HF}$) obtained in Secs.~\ref{subsec:cef_refinement} and \ref{subsec:hyperfine_refinement} were used in the calculations.  
The experimental data in panel (a) are plotted as $(\chi - \chi_0)^{-1}$.  
The inset in panel (c) shows an enlarged view of $M(H)$ below 2~T.}
\end{figure}

Figure~\ref{fig:chiT_and_MH_calc} compares the experimental results with calculations including the contribution of Ho nuclear spins to $\chi^{-1}(T)$, $\chi(T)$, and $M(H)$.  
The calculated curves, based on the refined parameters from Secs.~\ref{subsec:cef_refinement} and \ref{subsec:hyperfine_refinement}, reproduce the experimental data well.

%%%%%%%%%%%%%%%%%%%%%%%%%%%%%%%%%%%%%%%
%%%  Fig. 10
%%%%%%%%%%%%%%%%%%%%%%%%%%%%%%%%%%%%%%%
%%%%%%%%%%%%%%%%%%%%%%%%%%%%%%%%%%%%%%%

%
%
%
%%%%%%%%%%%%%%%%%%%%%%%%%%%%%%%%%%%%%%%
\section{\label{app_sec:C_LuCo2Zn20} Specific heat of \boldmath LuCo$_2$Zn$_{20}$}
%%%%%%%%%%%%%%%%%%%%%%%%%%%%%%%%%%%%%%%

\subsection{\label{app_subsec:C_LuCo2Zn20_results} Experimental results}

\begin{figure}[tb!]
\includegraphics[keepaspectratio, width=8cm, clip]{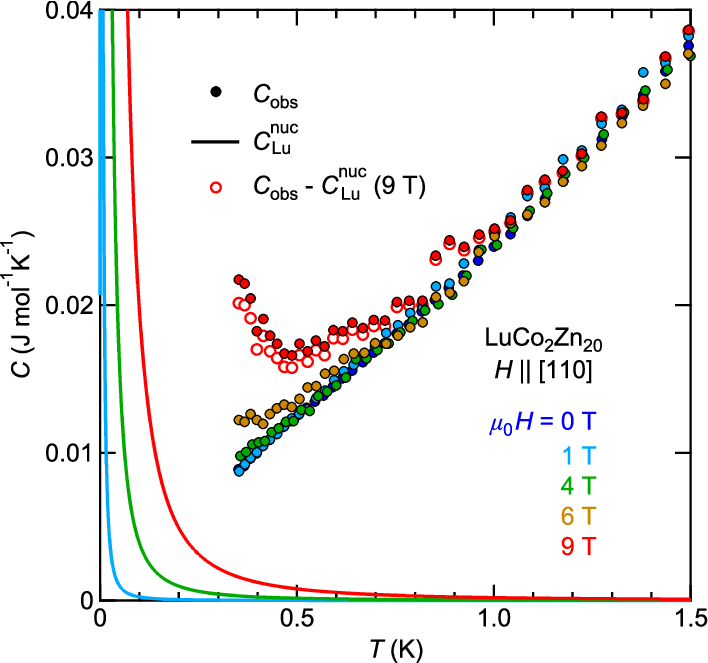}
\caption{\label{fig:C_LCZ} 
Temperature dependence of the specific heat in LuCo$_2$Zn$_{20}$, $C_\mathrm{obs}$ (closed circles), and the nuclear specific heat contribution of Lu calculated from Eq.~\eqref{eq:C_Lu_cal}, $C_\mathrm{Lu}^\mathrm{nuc}$ (solid lines). 
Open circles represent $C_\mathrm{obs} - C_\mathrm{Lu}^\mathrm{nuc}$ at $\mu_0 H = 9 \ \mathrm{T}$, corresponding to $C_\mathrm{nonmag} + C^\mathrm{nuc}_\mathrm{Co} + C^\mathrm{nuc}_\mathrm{Zn}$ at 9 T (see Appendix~\ref{app_subsec:C_bg} for details).}
\end{figure}

Figure~\ref{fig:C_LCZ} shows the temperature dependence of the observed specific heat, $C_\mathrm{obs}(T)$, in LuCo$_2$Zn$_{20}$, a nonmagnetic reference compound for HoCo$_2$Zn$_{20}$.  
The magnetic field was applied along [110], as in the specific heat measurements of HoCo$_2$Zn$_{20}$ shown in Fig.~\ref{fig:CT_and_rhoT}.  
Comparing $C_\mathrm{obs}$ at 0, 1, 4, 6, and 9 T at the same temperature, we find that $C_\mathrm{obs}$ increases with increasing field, and this increase becomes more pronounced at lower temperatures.  
These behaviors indicate the presence of nuclear specific heat arising from Lu, Co, and Zn isotopes listed in Table~\ref{tab:list_nuclear_spin}.  
However, comparison of the vertical scales in Figs.~\ref{fig:CT_and_rhoT}(a) and \ref{fig:C_LCZ} clearly shows that, in HoCo$_2$Zn$_{20}$, the nuclear contributions of Co and Zn, $C^\mathrm{nuc}_\mathrm{Co}$ and $C^\mathrm{nuc}_\mathrm{Zn}$, are much smaller than the specific heat contribution from Ho sites, $C_\mathrm{Ho}$, at least within the temperature range of the present study. 

%%%%%%%%%%%%%%%%%%%%%%%%%%%%%%%%%%%%%%%
%%%  Fig. 11
%%%%%%%%%%%%%%%%%%%%%%%%%%%%%%%%%%%%%%%
%%%%%%%%%%%%%%%%%%%%%%%%%%%%%%%%%%%%%%%

\subsection{\label{app_subsec:C_bg} Estimation of \boldmath $C_\mathrm{nonmag} + C^\mathrm{nuc}_\mathrm{Co} + C^\mathrm{nuc}_\mathrm{Zn}$}

Although $C^\mathrm{nuc}_\mathrm{Co} + C^\mathrm{nuc}_\mathrm{Zn}$ in Eq.~\eqref{eq:HCZ_C_component} is negligible in our study, 
$C_\mathrm{nonmag} + C^\mathrm{nuc}_\mathrm{Co} + C^\mathrm{nuc}_\mathrm{Zn}$ of HoCo$_2$Zn$_{20}$ can be estimated using the specific heat data of LuCo$_2$Zn$_{20}$. 
The observed specific heat of LuCo$_2$Zn$_{20}$, $C_\mathrm{obs}$, can be decomposed into four contributions:
\begin{equation}
    C_\mathrm{obs} = C_\mathrm{nonmag} + C^\mathrm{nuc}_\mathrm{Lu} + C^\mathrm{nuc}_\mathrm{Co} + C^\mathrm{nuc}_\mathrm{Zn},
    \label{eq:LCZ_C_component}
\end{equation}
where $C^\mathrm{nuc}_\mathrm{Lu}$ is the nuclear specific heat of Lu.  
To subtract $C^\mathrm{nuc}_\mathrm{Lu}$ from $C_\mathrm{obs}$, we analytically calculated $C^\mathrm{nuc}_\mathrm{Lu}$. 

Since Lu occupies a site with cubic symmetry, the nuclear quadrupole interaction at Lu sites can be neglected \cite{Abragam1961}.  
Thus, the degeneracy of the Lu nuclear spins is lifted only by the nuclear Zeeman effect. 
For an isotope of Lu with mass number $i$ ($=175,176$), the eigenenergy is $E^i = \mu_0 \mu_\mathrm{N} g_\mathrm{N}^i H I_z^i \ (I_z^i = -I^i, -I^i+1, \ldots, I^i)$, where $g_\mathrm{N}^i$ and $I^i$ are the nuclear $g$ factor and nuclear spin, respectively. 
The corresponding free energy is
\begin{equation}
F^i = - k_\mathrm{B} T \ln \sum_{I_z^i = -I^i}^{I^i} \exp(-\beta \mu_0 \mu_\mathrm{N} g_\mathrm{N}^i H I_z^i),
\end{equation}
where $k_\mathrm{B}$ is the Boltzmann constant and $\beta = (k_\mathrm{B}T)^{-1}$. 
The specific heat for each Lu isotope is then given by
\begin{equation}
c^i = - T \frac{\partial^2 F^i}{\partial T^2} 
     = - \varepsilon_I^i \frac{\partial B_I(\beta \varepsilon_I^i)}{\partial T},
\end{equation}
where $B_I(x)$ is the Brillouin function and $\varepsilon_I^i = \mu_0 \mu_\mathrm{N} g_\mathrm{N}^i H I^i$.  
Finally, using the natural abundance $n_a^i$ of Lu isotopes and the Avogadro constant $N_\mathrm{A}$, the molar nuclear specific heat of Lu is obtained as
\begin{equation}
    C^\mathrm{nuc}_\mathrm{Lu} = N_\mathrm{A} \left(n_a^{175} c^{175} + n_a^{176} c^{176}\right).
    \label{eq:C_Lu_cal}
\end{equation}

Based on $n_a^i$, $I^i$, and $g_\mathrm{N}^i$ listed in Table~\ref{tab:list_nuclear_spin}, we calculated $C^\mathrm{nuc}_\mathrm{Lu}$ at $\mu_0 H =$ 1, 4, and 9 T using Eq.~\eqref{eq:C_Lu_cal}.  
The results are shown as solid lines in Fig.~\ref{fig:C_LCZ}.  
We then subtracted $C^\mathrm{nuc}_\mathrm{Lu}$ from $C_\mathrm{obs}$ to obtain $C_\mathrm{nonmag} + C^\mathrm{nuc}_\mathrm{Co} + C^\mathrm{nuc}_\mathrm{Zn}$.  
Since the calculated $C^\mathrm{nuc}_\mathrm{Lu}$ at 1 and 4 T is nearly zero above 0.3 K, only the data for $C_\mathrm{obs} - C^\mathrm{nuc}_\mathrm{Lu}$ at 9 T are shown in Fig.~\ref{fig:C_LCZ}.

%
%
%
%%%%%%%%%%%%%%%%%%%%%%%%%%%%%%%%%%%%%%%
\section{\label{app_sec:analysis_method} Analysis method}
%%%%%%%%%%%%%%%%%%%%%%%%%%%%%%%%%%%%%%%

In this appendix, we define $\mathcal{H}_\alpha'$ as $\mathcal{H}_\alpha$ with the term $J_\mathrm{ex} \langle \bm{J}_\alpha \rangle \cdot \langle \bm{J}_\beta \rangle / 2$ in Eq.~\eqref{eq:Hamil_MF_single} removed.

\subsection{\label{app_subsec:analysis_method_1} Method in Sec. \ref{subsec:cef_refinement}}

In Sec.~\ref{subsec:cef_refinement}, the cubic CEF parameters $W$ and $x$, along with the magnetic exchange constant $J_\mathrm{ex}$, were refined by comparing the observed magnetization curves $M(H)$ at 2 and 10~K for $H \parallel [100]$, $[110]$, and $[111]$ (Fig.~\ref{fig:MH_2K10K}) with calculations based on a model without Ho nuclear spins.  
The six $M(H)$ curves, sampled every 0.5~T from 0.5 to 7~T, provided a total of 84 data points for the optimization.  
The magnetization per Ho site, neglecting nuclear spins, was calculated as
\begin{equation}
M_\mathrm{cal} = - \frac{\mu_\mathrm{B} g_J}{d} \, \bm{e}_H \cdot (\langle \bm{J}_A \rangle + \langle \bm{J}_B \rangle),
\end{equation}
where $\bm{e}_H = \bm{H}/H$ is the unit vector along $\bm{H}$ and $d$ is the number of sublattices ($d = 1$ for $J_\mathrm{ex} \ge 0$ and $d = 2$ for $J_\mathrm{ex} < 0$).  
For $J_\mathrm{ex} \ge 0$, the $\langle \bm{J}_B \rangle$ term was omitted.  
We refined $W$, $x$, and $J_\mathrm{ex}$ simultaneously by minimizing the sum of squared deviations $(M_\mathrm{cal} - M)^2$ over 84 data points.  
To locate the global optimum within $|W| \leq 0.2$~K, $|x| \leq 1$, and $|J_\mathrm{ex}| \leq 0.2$~K, we employed the JADE algorithm~\cite{Zhang2009}, an improved variant of differential evolution~\cite{Storn1997}.  
The method for evaluating $\langle \bm{J}_A \rangle$ and $\langle \bm{J}_B \rangle$ under trial values of $W$, $x$, and $J_\mathrm{ex}$ generated by the JADE algorithm is described below.

To solve the self-consistent total Hamiltonian of Eq.~\eqref{eq:Hamil_MF_total} for both $J_\mathrm{ex} \ge 0$ and $J_\mathrm{ex} < 0$, we searched for the $x$, $y$, and $z$ components of $\langle \bm{J}_A \rangle$ that minimize the free energy per Ho site:
\begin{equation}
F = 
    \begin{dcases}
        -\frac{1}{\beta} \ln Z_A + \frac{J_\mathrm{ex}}{2} \langle \bm{J}_A \rangle^2 & (J_\mathrm{ex} \ge 0), \\
        -\frac{1}{2\beta} \ln(Z_A Z_B) + \frac{J_\mathrm{ex}}{2} \langle \bm{J}_A \rangle \cdot \langle \bm{J}_B \rangle & (J_\mathrm{ex} < 0),
    \end{dcases}
\label{eq:total_free_energy}
\end{equation}
where $Z_i = \sum_n \exp(-\beta \varepsilon_n^i)$ ($i = A, B$; $\varepsilon_n^i$: eigenenergy of $\mathcal{H}'_i$) is the partition function of $\mathcal{H}'_i$.  
In the case of $J_\mathrm{ex} < 0$, $\langle \bm{J}_B \rangle = \mathrm{Tr}(\bm{J}_B \, e^{-\beta \mathcal{H}'_B})/Z_B$ depends on $\langle \bm{J}_A \rangle$ because $\mathcal{H}'_B$ contains $\langle \bm{J}_A \rangle$.  
The optimal $\langle \bm{J}_A \rangle$ was obtained using the JADE algorithm under the constraint $|\langle \bm{J}_A \rangle| \leq J \ (= 8)$.  
For $J_\mathrm{ex} < 0$, the procedure was as follows:  
(1) substitute a trial $\langle \bm{J}_A \rangle$ generated by JADE into $\mathcal{H}_B'$, and solve for the eigenenergies $\varepsilon_n^B$ and eigenfunctions $\ket{n_B}$ of $\mathcal{H}'_B$;  
(2) calculate $\langle \bm{J}_B \rangle$ via
\begin{equation}
    \langle \bm{J}_B \rangle = \frac{1}{Z_B} \sum_n \bra{n_B} \bm{J}_B \ket{n_B} \exp(-\beta \varepsilon_n^B) \,;
    \label{eq:J_B_canonical}
\end{equation} 
(3) substitute $\langle \bm{J}_B \rangle$ into $\mathcal{H}_A'$, and solve for the eigenenergies $\varepsilon_n^A$ of $\mathcal{H}_A'$;  
(4) evaluate the free energy $F$ using Eq.~\eqref{eq:total_free_energy}; 
and (5) repeat steps (1)–(4) until convergence within the JADE algorithm.  
For $J_\mathrm{ex} \ge 0$, step (1) was applied with the suffix “$B$” replaced by “$A$”, and steps (2) and (3) were omitted.  
Because $J = 8$ and nuclear spins were neglected, the dimension of each $\mathcal{H}'_\alpha$ matrix was 17.

With the refined parameters $W = 0.0443$~K, $x = -0.0640$, and $J_\mathrm{ex} = -0.0511$~K, we then calculated the specific heat contribution from Ho sites, $C_\mathrm{Ho}(T)$, shown in Fig.~\ref{fig:C_comparison}(b).  
Since the entropy of Ho sites is given by $S(T) = \int_0^T C_\mathrm{Ho}(T)/T \, dT$, the calculated $C_\mathrm{Ho}(T)$ was obtained from
\begin{equation}
C_\mathrm{Ho}^\mathrm{cal}(T) = T \frac{S(T+\Delta T) - S(T-\Delta T)}{2 \Delta T},
\label{eq:C_Ho^cal}
\end{equation}
with a temperature increment $\Delta T = 0.5$~mK.  
Here, $S(T \pm \Delta T)$ was evaluated using the calculated $F$ from Eq.~\eqref{eq:total_free_energy} and the thermal average of the energy per Ho site,
\begin{equation}
E = \frac{1}{2} \sum_{i = A, B} \sum_n \frac{\varepsilon_n^i e^{-\beta \varepsilon_n^i}}{Z_i},
\end{equation}
via the relation $S(T) = (E(T) - F(T))/T$.

\subsection{\label{app_subsec:analysis_method_2} Method in Sec. \ref{subsec:hyperfine_refinement}}

In Sec.~\ref{subsec:hyperfine_refinement}, the hyperfine coupling constant $A_\mathrm{HF}$ was refined by comparing the specific heat $C_\mathrm{Ho}(T)$ below 3~K at 1, 4, and 9~T.  
As shown by the open circles in Fig.~\ref{fig:C_comparison}(a), five equally spaced interpolated points of $C_\mathrm{Ho}(T)$ on a logarithmic temperature scale between 0.38 and 3~K were used for this refinement.  
We optimized $A_\mathrm{HF}$ with the JADE algorithm by minimizing the sum of squared deviations $(C_\mathrm{Ho}^\mathrm{cal} - C_\mathrm{Ho})^2$ over 15 data points, where $C_\mathrm{Ho}^\mathrm{cal}$ was calculated using Eq.~\eqref{eq:C_Ho^cal}.  
The search range was set not to $|A_\mathrm{HF}| \leq 2000$~MHz (0.096~K) but to $0 \leq A_\mathrm{HF} \leq 2000$~MHz, since $A_\mathrm{HF} = 23 g_\mathrm{N} \mu_\mathrm{N} \mu_\mathrm{B} \braket{r^{-3}}/15$ \cite{Wybourne_note, Wybourne1965} is positive, as mentioned in Sec.~\ref{subsec:level_scheme}.
In these calculations, the refined values of $W$, $x$, and $J_\mathrm{ex}$ obtained in Appendix~\ref{app_subsec:analysis_method_1} were fixed.  
Unlike Appendix~\ref{app_subsec:analysis_method_1}, the nuclear Zeeman term $\mu_0 g_\mathrm{N} \mu_\mathrm{N} \bm{I}_\alpha \cdot \bm{H}$ and the hyperfine interaction term $A_\mathrm{HF} \bm{I}_\alpha \cdot \bm{J}_\alpha$ were included in $\mathcal{H}'_\alpha$.  
The solution of the total Hamiltonian [Eq.~\eqref{eq:Hamil_MF_total}] followed the procedure described in the second paragraph of Appendix~\ref{app_subsec:analysis_method_1}.  
Since the system considered consists of the $I = 7/2$ nuclear spin coupled with $f$ electrons of $J = 8$, the dimension of each $\mathcal{H}'_\alpha$ matrix was $8 \times 17 = 136$.

With the refined value $A_\mathrm{HF} = 0.0355$~K, we calculated the specific heat and magnetization, shown in Figs.~\ref{fig:C_comparison}(c) and \ref{fig:chiT_and_MH_calc}, respectively.  
The magnetization per Ho site was given by
\begin{equation}
M_\mathrm{cal} = -\frac{\mu_\mathrm{B}}{2} \, \bm{e}_H \cdot \sum_{i = A, B} \left( g_J \langle \bm{J}_i \rangle + g_\mathrm{N} \langle \bm{I}_i \rangle \right),
\end{equation}
where $\langle \bm{I}_i \rangle$ is the thermal average of $\bm{I}_i$.  
Finally, in the $H-T$ phase diagram of Fig.~\ref{fig:phase_diagram}, we plotted the transition temperature at which $\langle \bm{J}_A \rangle = \langle \bm{J}_B \rangle$ changes to $\langle \bm{J}_A \rangle \neq \langle \bm{J}_B \rangle$.

%%%%%%%%%%%%%%%%%%%%%%%%%%%%%%%%%%%%%%%
\section{\label{app_sec:splitting_CEF} Splitting of the CEF ground level due to hyperfine coupling} 
%%%%%%%%%%%%%%%%%%%%%%%%%%%%%%%%%%%%%%%

In this Appendix, we explain how to construct the energy diagram shown in Fig.~\ref{fig:HCZ_level_scheme}(b).  
When the cubic CEF parameter $W$ is positive, the CEF ground state for $J = 8$ can be classified into four cases depending on the other cubic CEF parameter $x$:  
(1) a $\Gamma_1$ singlet, 
(2) a $\Gamma_5$ triplet, 
(3) a direct sum of a $\Gamma_1$ singlet and a $\Gamma_5$ triplet, and  
(4) a direct sum of a $\Gamma_3$ doublet and a $\Gamma_5$ triplet.  
These cases are realized for (1) $-1 \leq x < -38/83$, (2) $-38/83 < x \leq 1$ with $x \neq 2/3$, (3) $x = -38/83$, and (4) $x = 2/3$, respectively. 
The $f$-electron wave functions for the $\Gamma_1$, $\Gamma_3$, and $\Gamma_5$ CEF states can be expressed in terms of the $J_z$ components~\cite{Lea1962}:  
\begin{align}
    \Ket{\Gamma_1}	=& \frac{\sqrt{390}}{48} \left( \left| 8 \right\rangle + \left| -8 \right\rangle \right) 
                     + \frac{\sqrt{42}}{24} \left( \left| 4 \right\rangle + \left| -4 \right\rangle \right) \notag \\
                     & + \frac{\sqrt{33}}{8} \left| 0 \right\rangle, \\
    \Ket{\Gamma_3 ; \alpha}	=& a_1(x) \left( \left| 8 \right\rangle + \left| -8 \right\rangle \right) 
                               + a_2(x) \left( \left| 4 \right\rangle + \left| -4 \right\rangle \right) \notag \\ 
                               & + a_3(x) \left| 0 \right\rangle, \\
    \Ket{\Gamma_3 ; \beta}	=& b_1(x) \left( \left| 6 \right\rangle + \left| -6 \right\rangle \right) 
                               + b_2(x) \left( \left| 2 \right\rangle + \left| -2 \right\rangle \right), \\
    \Ket{\Gamma_5 ; \pm}	=& c_1(x) \left|\pm 7\right\rangle + c_2(x) \left|\pm 3\right\rangle \notag \\
                               & + c_3(x) \left|\mp 1\right\rangle + c_4(x) \left|\mp 5\right\rangle, \\
    \Ket{\Gamma_5 ; 0}	=& d_1(x) \left( \left| 6 \right\rangle - \left| -6 \right\rangle \right) 
                               + d_2(x) \left( \left| 2 \right\rangle - \left| -2 \right\rangle \right),
\end{align}
where the coefficients $a_i$, $b_i$, $c_i$, and $d_i$ depend on $x$. 
Below, we consider a composite system consisting of the nuclear spin space $I = 7/2$ of the $^{165}$Ho isotope and the CEF ground subspace corresponding to each of the above cases, and describe how the ground multiplet is split by the hyperfine coupling $A_\mathrm{HF}\,\bm{I}\cdot\bm{J}$.  

% (i) $\Gamma_1$ case
(1) $\Gamma_1$ \textit{subspace:} Since the magnetic dipole moment is inactive in the $\Gamma_1$ state, the $f$ electrons do not couple to the nuclear spin via hyperfine interaction.  
Therefore, the eightfold nuclear-spin multiplet remains unsplit, and the eigenfunctions are simply given by the product states $\Ket{\Gamma_1}\Ket{I_z}$ with $I_z = -7/2, -5/2, \dots, 7/2$.  

% (ii) $\Gamma_5$ subspace
(2) $\Gamma_5$ \textit{subspace:} We numerically calculated the matrix elements of $J_z$, the raising operator $J_+ (= J_x + i J_y)$, and the lowering operator $J_- (= J_x - i J_y)$ in the $\Gamma_5$ state space.  
These operators share the same coefficient $\alpha(x)$:
\begin{equation}
J_z = \alpha(x)
\begin{blockarray}{rccc}
 & \Ket{\Gamma_5 ; +} & \Ket{\Gamma_5 ; 0} & \Ket{\Gamma_5 ; -} \\
\begin{block}{r@{\quad}(ccc)}
  \Bra{\Gamma_5 ; +} & 1 & 0 & 0  \\
  \Bra{\Gamma_5 ; 0} & 0 & 0 & 0 \\
  \Bra{\Gamma_5 ; -} & 0 & 0 & -1 \\
\end{block}
\end{blockarray}
\label{eq:G5_J_z_matrix}
\end{equation}
\begin{equation}
J_+ = J_-^\dagger = \alpha(x)
\begin{blockarray}{rccc}
 & \Ket{\Gamma_5 ; +} & \Ket{\Gamma_5 ; 0} & \Ket{\Gamma_5 ; -} \\
\begin{block}{r@{\quad}(ccc)}
  \Bra{\Gamma_5 ; +} & 0 & \sqrt{2} & 0  \\
  \Bra{\Gamma_5 ; 0} & 0 & 0 & \sqrt{2} \\
  \Bra{\Gamma_5 ; -} & 0 & 0 & 0 \\
\end{block}
\end{blockarray}
\label{eq:G5_J_pm_matrix}
\end{equation}
The $3 \times 3$ matrices in Eqs.~\eqref{eq:G5_J_z_matrix} and \eqref{eq:G5_J_pm_matrix} correspond to the $z$ component $S_z$ and the raising operator $S_+$, respectively, in the $S = 1$ spin space.  
Thus, $\bm{J}$ can be expressed as $\bm{J} = \alpha(x)\,\bm{S}$.  
The numerically obtained values of $\alpha(x)$ are plotted in Fig.~\ref{fig:HCZ_level_scheme}(b) as the light-purple line.

By coupling the effective spin $S = 1$ with the nuclear spin $I = 7/2$ via the hyperfine interaction $A_\mathrm{HF} \alpha(x) \bm{I} \cdot \bm{S}$, the 24-fold multiplet splits into a sextet, an octet, and a dectet.  
As stated in Sec.~\ref{subsec:level_scheme}, each multiplet is characterized by the new total angular momentum $\bm{F} = \bm{I} + \bm{S}$.  
We calculated the eigenenergies for each multiplet using Eq.~\eqref{eq:HF_eigenenergy}, and plotted the $F = 5/2$ sextet, $F = 7/2$ octet, and $F = 9/2$ dectet in Fig.~\ref{fig:HCZ_level_scheme}(b) as blue, green, and orange lines, respectively. 
We note that the wave functions of each multiplet can be expressed using the Clebsch-Gordan coefficients $\braket{I = 7/2, I_z, S = 1, S_z | F, F_z}$. 
For example, the $F_z = \pm 1/2, \pm 3/2, \pm 5/2$ wave functions of the $F = 5/2$ sextet can be written as
\begin{align} 
&\Ket{F_z = \pm 1/2} = \notag \\ & \quad \sqrt{\frac{5}{14}} \ket{\pm 3/2, \mp} - \sqrt{\frac{3}{7}} \ket{\pm 1/2, 0} + \sqrt{\frac{3}{14}} \ket{\mp 1/2, \pm}, \label{eq:Fz_1} \\
&\Ket{F_z = \pm 3/2} = \notag \\
& \quad \frac{1}{2}\sqrt{\frac{15}{7}} \ket{\pm 5/2, \mp} - \sqrt{\frac{5}{14}} \ket{\pm 3/2, 0} + \frac{1}{2}\sqrt{\frac{3}{7}} \ket{\pm 1/2, \pm}, \label{eq:Fz_2} \\
&\Ket{F_z = \pm 5/2} = \notag \\
& \quad \frac{\sqrt{3}}{2} \ket{\pm 7/2, \mp} - \sqrt{\frac{3}{14}} \ket{\pm 5/2, 0} + \frac{1}{2\sqrt{7}} \ket{\pm 3/2, \pm}, \label{eq:Fz_3} 
\end{align}
where the basis states on the right-hand side of Eqs.~\eqref{eq:Fz_1}–\eqref{eq:Fz_3}, $\ket{I_z, S_z}$, denote the product states of the nuclear spin with $I_z$ and the $f$-electron effective spin with $S_z$ ($S_z = \pm, 0$ correspond to $\ket{\Gamma_5; \pm}$ and $\ket{\Gamma_5; 0}$, respectively).

% (iii) $\Gamma_1 \oplus \Gamma_5$ subspace
(3) $\Gamma_1 \oplus \Gamma_5$ \textit{subspace:}  
The coefficients of the $\Gamma_5$ wave functions, $c_i(x)$ and $d_i(x)$, for $x = -38/83$ are given by
\begin{gather}
    \begin{aligned}
    &(c_1, c_2, c_3, c_4) \\
    & = \frac{C}{64} \left(95\sqrt{65}, 59\sqrt{105}, -113\sqrt{11}, -13\sqrt{91} \right),
    \end{aligned} \\
    (d_1, d_2) = C \left(\sqrt{39}, \frac{\sqrt{385}}{2} \right),
\end{gather}
with $C = \sqrt{2/541}$.  
At this parameter, the matrices of $J_z$ and $J_\pm$ in the $\Gamma_1 \oplus \Gamma_5$ subspace are
\begin{equation}
J_z =
\begin{blockarray}{rcccc}
 & \Ket{\Gamma_1} & \Ket{\Gamma_5 ; +} & \Ket{\Gamma_5 ; 0} & \Ket{\Gamma_5 ; -} \\
\begin{block}{r@{\quad}(c|ccc)}
  \Bra{\Gamma_1} & 0 & 0 & 0 & 0 \mathstrut \\ \cline{2-5} 
  \Bra{\Gamma_5 ; +} & 0 & \alpha' \rule{0pt}{2.5ex} & 0 & 0  \\
  \Bra{\Gamma_5 ; 0} & 0 & 0 & 0 & 0 \\
  \Bra{\Gamma_5 ; -} & 0 & 0 & 0 & -\alpha' \\
\end{block}
\end{blockarray}
\label{eq:G1G5_Jz}
\end{equation}
and
\begin{equation}
J_+ = J_-^\dagger =
\begin{blockarray}{rcccc}
 & \Ket{\Gamma_1} & \Ket{\Gamma_5 ; +} & \Ket{\Gamma_5 ; 0} & \Ket{\Gamma_5 ; -} \\
\begin{block}{r@{\quad}(c|ccc)}
  \Bra{\Gamma_1} & 0 & 0 & 0 & 0 \mathstrut \\ \cline{2-5} 
  \Bra{\Gamma_5 ; +} & 0 & 0 & \alpha' \sqrt{2} \rule{0pt}{2.5ex} & 0  \\
  \Bra{\Gamma_5 ; 0} & 0 & 0 & 0 & \alpha' \sqrt{2} \\
  \Bra{\Gamma_5 ; -} & 0 & 0 & 0 & 0 \\
\end{block}
\end{blockarray}
\ \ ,
\label{eq:G1G5_Jpm}
\end{equation}
with $\alpha' = 77 \ 899/17 \ 312 \approx 4.50$.  
Because the off-diagonal block matrices in Eqs.~\eqref{eq:G1G5_Jz} and \eqref{eq:G1G5_Jpm} vanish, the $\Gamma_1$ and $\Gamma_5$ subspaces can be treated independently.  
Therefore, the composite system reduces to two independent cases: the coupling of the $I = 7/2$ nuclear spin with the $\Gamma_1$ state [case (1)] and with the $\Gamma_5$ state [case (2)].  
As a result, the 32-fold manifold splits into $F = 5/2$, $7/2$, and $9/2$ multiplets, together with a noninteracting octet.  
The eigenenergies of these $F$ multiplets, obtained from Eq.~\eqref{eq:HF_eigenenergy} by substituting $\alpha = \alpha'$, and the eigenenergy of the noninteracting octet are plotted as closed circles in Fig.~\ref{fig:HCZ_level_scheme}(b). 

% (iv) $\Gamma_3 \oplus \Gamma_5$ subspace
(4) $\Gamma_3 \oplus \Gamma_5$ \textit{subspace:}  
The coefficients of the $\Gamma_3$ and $\Gamma_5$ wave functions for $x = 2/3$ are
\begin{gather}
    (a_1, a_2, a_3) = \frac{1}{16} \left(\frac{\sqrt{5}}{2}, \sqrt{91}, -\frac{\sqrt{286}}{2} \right), \\
    (b_1, b_2) = (d_1, d_2) = \left(\frac{1}{\sqrt{2}}, 0 \right), \\
    (c_1, c_2, c_3, c_4) = \frac{1}{32\sqrt{2}} \left(3\sqrt{15}, \sqrt{455}, -\sqrt{429}, 7\sqrt{21} \right).
\end{gather}
In contrast to case (3), the off-diagonal block matrices of $J_z$ and $J_+ (= J_-^\dagger)$ contain nonzero elements:
\begin{gather}
    \Bra{\Gamma_3; \alpha} J_z \Ket{\Gamma_5; 0} = 6, \\
    \Bra{\Gamma_3; \alpha} J_+ \Ket{\Gamma_5; -} = \Bra{\Gamma_5; +} J_+ \Ket{\Gamma_3; \alpha} = 3 \sqrt{2}, \\
    \Bra{\Gamma_3; \beta} J_+ \Ket{\Gamma_5; +} = \Bra{\Gamma_5; -} J_+ \Ket{\Gamma_3; \beta} = 3 \sqrt{6}.
\end{gather}
Therefore, we must consider the composite state formed by coupling the $I = 7/2$ nuclear spin with the direct-sum space $\Gamma_3 \oplus \Gamma_5$.  
Numerical diagonalization of the $A_\mathrm{HF} \bm{I} \cdot \bm{J}$ matrix of size $40 \times 40$ shows that the 40-fold manifold splits into seven quartets and six doublets.  
The eigenenergies, shown as triangles in Fig.~\ref{fig:HCZ_level_scheme}(b), are $\varepsilon/A_\mathrm{HF} =$ $-24.7$ (4), $-19.4$ (2), $-17.5$ (4), $-10.6$ (2), $-8.2$ (4), $-4.9$ (2), $2.5$ (2), $2.9$ (4), $5.5$ (4), $16.9$ (2), $18.1$ (4), $20.4$ (2), and $21.6$ (4), where the numbers in parentheses indicate degeneracy.

%%%%%%%%%%%%%%%%%%%%%%%%%%%%%%%%%%%%%%%
\section{\label{app_sec:residual_entropy} Experimental proposal for capturing the residual entropy} 
%%%%%%%%%%%%%%%%%%%%%%%%%%%%%%%%%%%%%%%

In this Appendix, we outline an experimental strategy for capturing the residual entropy, which is one of the most compelling experimental signatures of the MCK effect, in diluted cubic systems containing nuclear spins at magnetic sites.
At 0~T, the MCK effect may occur because the electron-nuclear spin degrees of freedom remain active owing to the degeneracy of the hyperfine-coupled states.
In contrast, under finite magnetic fields, the MCK effect is not expected, since the Zeeman splitting removes the degeneracy of these states.
In this regime, the Hamiltonian for each magnetic site can be described by a single-site model consisting of the cubic CEF Hamiltonian, the Zeeman terms for the electronic and nuclear moments, and the hyperfine interaction:
\begin{equation}
    \mathcal{H}_\mathrm{single} =
    \mathcal{H}_\mathrm{CEF}^\mathrm{cubic}
    + \mu_0 \left( g_J \mu_\mathrm{B} \bm{J}
    + g_\mathrm{N} \mu_\mathrm{N} \bm{I} \right) \cdot \bm{H}
    + A_\mathrm{HF} \bm{I} \cdot \bm{J}.
    \label{eq:Hamil_single}
\end{equation}

The experimental procedure for estimating the residual entropy at 0~T is as follows.
(1) First, the cubic CEF parameters and the hyperfine coupling constant $A_\mathrm{HF}$ in Eq.~\eqref{eq:Hamil_single} are determined from experiments performed under finite magnetic fields.
As shown in Secs.~\ref{subsec:cef_refinement} and \ref{subsec:hyperfine_refinement}, these parameters can be obtained from magnetization and specific heat measurements in magnetic fields.
(2) Next, isothermal magnetization measurements are performed as a function of magnetic field from $B = 0$ to $B'$, where $B = \mu_0 H$, at two nearby temperatures, $T = T' + \Delta T$ and $T = T' - \Delta T$.
Here, $T'$ is a fixed temperature and $\Delta T$ is a small temperature interval.
$B'$ is defined as a magnetic field such that the Hamiltonian at magnetic sites can be well described by Eq.~\eqref{eq:Hamil_single}.
Accordingly, $B'$ is chosen so that the experimentally measured thermodynamic quantities quantitatively agree with those calculated from Eq.~\eqref{eq:Hamil_single} using the refined parameters.
The entropy at $T = T'$ and $B = 0$ can then be estimated using Maxwell's relation $(\partial S/\partial B)_T = (\partial M/\partial T)_B$:
\begin{align}
S(T', B = 0) &= S(T', B') - \int_0^{B'} \left( \frac{\partial M}{\partial T} \right)_B dB
\label{eq:S_0T} \\
&\approx S(T', B') - \frac{1}{2\Delta T} \int_0^{B'} \Delta M \, dB,
\end{align}
where
\begin{equation}
\Delta M = M(T' + \Delta T, B) - M(T' - \Delta T, B),
\end{equation}
and the entropy $S(T', B')$ is calculated from Eq.~\eqref{eq:Hamil_single} using the refined parameters.

As an alternative to isothermal magnetization measurements, the quantity $(\partial M/\partial T)_B$ in Eq.~\eqref{eq:S_0T} can also be obtained from measurements of the magnetic Gr\"{u}neisen ratio $\Gamma_\mathrm{H}$ under adiabatic conditions~\cite{Tokiwa2011} and the specific heat from $B = 0$ to $B'$ at $T = T'$, using the thermodynamic relation $(\partial M/\partial T)_B = - C_\mathrm{m} \Gamma_\mathrm{H}$~\cite{Tokiwa2011}.
Here, $C_\mathrm{m}$ denotes the specific heat contribution from magnetic sites.
In the case of diluted Ho systems, $C_\mathrm{m}$ corresponds to $C_\mathrm{Ho}$ in this paper.

(3) Finally, the temperature dependence of the specific heat at $B = 0$ is measured down to the lowest accessible temperatures, and the entropy at $B = 0$ is evaluated via $S(T) = S(T') - \int_T^{T'} C_\mathrm{m}(T)/T \, dT$.
If $S(T)$ asymptotically approaches a finite positive value as $T \to 0$, the presence of residual entropy can be established.
We note that $T'$ is, in principle, an arbitrary finite temperature; however, in practice, it should be chosen below a few kelvin, where $C_\mathrm{m} (T)$ can be accurately estimated.

% The \nocite command causes all entries in a bibliography to be printed out
% whether or not they are actually referenced in the text. This is appropriate
% for the sample file to show the different styles of references, but authors
% most likely will not want to use it.
%\nocite{*}

\bibliography{bibliography}% Produces the bibliography via BibTeX.

\end{document}